\pdfoutput=1
\documentclass[useAMS,british]{tRPC2e}

\usepackage{amsmath,amssymb}
\usepackage[T1]{fontenc}
\usepackage[utf8]{inputenc}
\usepackage{xspace}
\usepackage{hyperref}
\usepackage[todos]{CMImacros}
\presetkeys{todonotes}{size=\footnotesize}{}
\newcommand{\cis}{\emph{cis}\xspace}
\newcommand{\trans}{\emph{trans}\xspace}
\newcommand{\indw}{indole(H$_2$O)\xspace}%
\newcommand{\indww}[1]{indole(H$_2$O)$_{#1}$\xspace}%
\newcommand{\ortho}{\emph{ortho}\xspace}%
\newcommand{\para}{\emph{para}\xspace}%
\begin{document}
\setcounter{page}{1}%
\doi{10.1080/0144235X.YYYY.CATSid}%
\issn{1366--591X}%
\issnp{0144--235X} \jvol{00} \jnum{00} \jyear{2015} \jmonth{Some--Other}%
\markboth{Y.\,P.~Chang, D.A.~Horke, S.~Trippel, and J.~Küpper}{%
   International Reviews in Physical Chemistry}%
\articletype{Review}%
\title{Spatially-controlled complex molecules and their applications}%
\author{Yuan-Pin Chang$^{{\rm a}\dagger}$, Daniel A.\ Horke$^{{\rm a}\dagger}$, Sebastian
   Trippel$^{{\rm a}\dagger}$ %
   and %
   Jochen Küpper$^{\rm a,b,c}$$^{\ast}$ \\[12pt]%
   \affilfont{%
      $^{\rm a}${\em{Center for Free-Electron Laser Science, DESY, \\
            Notkestrasse 85, 22607 Hamburg, Germany}};
      \\[3pt]
      $^{\rm b}${\em{Department of Physics, Universität Hamburg, \\
            Luruper Chaussee 149, 22761 Hamburg, Germany}}
      \\[3pt]
      $^{\rm c}${\em{The Hamburg Center for Ultrafast Imaging, Universität Hamburg, \\
            Luruper Chaussee 149, 22761 Hamburg, Germany}} }%
   \\[6pt]
   \received{\today}%
}%
\thanks{$^\ast$ E-mail: \url{jochen.kuepper@cfel.de}; website: \url{http://desy.cfel.de/cid/cmi} \\
   $^\dagger$ These authors contributed equally to this work.}
\maketitle
\begin{abstract}
   The understanding of molecular structure and function is at the very heart of the chemical and
   molecular sciences. Experiments that allow for the creation of structurally pure samples and the
   investigation of their molecular dynamics and chemical function have developed tremendously over
   the last few decades, although ``there's plenty of room at the bottom'' for better control as
   well as further applications.

   Here, we describe the use of inhomogeneous electric fields for the manipulation of neutral
   molecules in the gas-phase, \ie, for the separation of complex molecules according to size,
   structural isomer, and quantum state. For these complex molecules, all quantum states are
   strong-field seeking, requiring dynamic fields for their confinement. Current applications of
   these controlled samples are summarised and interesting future applications discussed.
\end{abstract}


\section{Introduction}
\label{sec:introduction}
Structure defines function. This structure-function relationship is a fundamental concept in the
molecular sciences. Chemistry textbooks define structure through atomic coordinates and bond orders,
using representations such as ball-and-stick models or Lewis formulas. In biology sometimes further
abstractions are made, although often as additional information to the nuclear geometry: the protein
data bank (PDB), for instance, typically contains coordinates of all atoms plus information on
folding motifs. At the same time, we realise that it is the electrons that form and break bonds and
that define the chemistry. Therefore, aren't it the electronic wavefunctions, or the electron
density, which define structure, and, thus function?

More generally, first we have to ask ourselves: Do molecules, in general, have a well-defined
structure? The answer to this question strongly depends on the actual definitions of a ``molecule''
and a ``structure'', that is, on the scientific background, or community, and the experiment
performed. In this work, we are concerned with isolated molecules in the gas-phase. The geometric
structures considered in chemistry and biology can typically be measured using spectroscopic
techniques~\cite{Kraitchman:AJP21:17, Pratt:ARPC49:481, Wuethrich:ACIE42:3340,
   Shipman:NewTechMW:2011, Perez:Science336:897} or diffractive-imaging
methods~\cite{Barty:ARPC64:415, Miller:Science343:1108}. We recognize that these structures depend,
for instance, on the electronic state of the molecule, but still this can be disentangled
spectroscopically even for large molecules~\cite{Ratzer:CP283:153}. Electronic structure can be
experimentally observed through the (multipole) moments of the charge distribution, such as precise
determinations of dipole moment vectors or polarizability tensors, possibly for individual
electronic states~\cite{deLeeuw:CPL7:288, Okruss:JCP110:10393}, but also coherent x-ray diffractive
imaging measures, conceptionally, electron density.

Here, we set out to take a more specific approach to ``molecular structure'', which is based on
interaction properties of the molecular species, for instance, with electromagnetic fields --
including these provided by other particles. Following this concept, \para and \ortho hydrogen
(H$_2$) are separate species, as they exhibit very different magnetic properties and they, as
isolated species, do not interconvert on typical experimental timescales. Equivalently, the
structural isomers of molecules in the gas-phase typically have to be considered different molecular
species, including the configurations of chiral molecules (enantiomers) or, at least in many cases,
the rotational isomers of complex molecules (conformers). Chemical functionality is, for instance,
intricately linked to this 3D geometric arrangement of molecules in space. This delicate
structure-function relationship is especially evident in many biological systems and determines,
\eg, protein folding and synthesis~\cite{Royer:ChemRev106:1769}, the conformations of sugars in
enzyme catalysis~\cite{Mayes:JACS136:1008}, or reactive pathways and
intermediates~\cite{Baldwin:NatChemBiol5:808}. Biology went to the extreme of homochirality and
organisms use only a single configuration of all molecules involved in
life~\cite{Blackmond:CSHPB2:a002147, Pross:Life}, which is a most-important structure-function
relationship that still awaits a conclusive explanation. We point out that this concept of molecular
structure is simply a description of the fact that, in quantum mechanics, the structure of a
molecule is given by the square of the wavefunction, its probability density. In order to follow a
reductionist, bottom-up description of molecular structure and function, we need to study and
understand the physical and chemical properties of these individual species, how they interconvert,
and how they specifically interact with their environment. This requires new experimental approaches
and techniques for the preparation and investigation of the probed samples, some of which are
detailed and discussed in this review.

When we prepare these molecules, or simply order them from the chemicals suppliers, we get
containers containing various ``molecules'', or structures, all for the price of one. A gas bottle
of hydrogen contains both, \para and \ortho H$_2$, and for glycine and tryptophan and many other
complex molecules we get multiple conformers in that one bottle~\cite{Suenram:JACS102:7180,
   Rizzo:JCP83:4819}. This is due to the fact that the interconversion between structural isomers is
rapid under the typical preparation and storage conditions, \eg, room temperature. Under such
conditions, these species, \ie, structural isomers, are simply well connected volumes of structural
phase space of these ``molecules''. The molecules rapidly explore the whole phase-space volume, \ie,
they change their conformation, that is, they change their ``text-book structure''.

In molecular beam experiments we produce samples of cold, isolated molecules using supersonic
expansions~\cite{Scoles:MolBeam:1and2, Fenn:ARPC47:1}, where the molecules are coexpanded into high
vacuum seeded in a high-pressure rare gas. In these beams the molecules are well isolated and the
rare remaining collision events are very cold, \ie, have a very small energy
impact~\cite{Erlekam:PCCP9:3786}. The molecules still interact with the surrounding black-body
radiation, but the typical timescales of the corresponding state-changing interactions are on the
order of seconds~\cite{Meerakker:PRL95:013003, Kuepper:FD142:155}, which is much longer than the
few-millisecond durations of the experiments.

\subsection{Manipulating molecular species with external fields}
The isolated molecules in these beams can be manipulated with external fields: Using electric
fields, they can be deflected, focused, and accelerated or decelerated~\cite{Meerakker:NatPhys4:595,
   Kuepper:FD142:155, Schnell:ACIE48:6010, Meerakker:CR112:2012}, they can be trapped in
space~\cite{Bethlem:Nature406:491, Meerakker:CR112:2012}, and they can be aligned or oriented, \ie,
fixed in space with respect to the directions of their dipole moments or polarizability
tensors~\cite{Loesch:JCP93:4779, Friedrich:PRL74:4623, Stapelfeldt:RMP75:543,
   Holmegaard:PRL102:023001}. These methods date back to the description of polar molecules in
inhomogeneous electric fields~\cite{Kallmann:ZP6:352} and Otto Stern's groundbreaking deflection
experiments of atoms in magnetic fields~\cite{Gerlach:ZP9:349} and of molecules in electric
fields~\cite{Wrede:ZP44:261}; the latter already being performed in Stern's institute for physical
chemistry in Hamburg -- with the original building nowadays being the historic home to our physics
department. From these times, magnetic and electric fields were used interchangeably, as
appropriate. Deceleration with time-varying magnetic fields has also been demonstrated during the
last decade~\cite{Vanhaecke:PRA75:031402, Narevicius:CR112:4879}.

Almost a century ago, Stern's group had not only deflected molecules with electric fields and
measured the electric dipole moment of potassium iodide~\cite{Wrede:ZP44:261}. Otto Stern also
realized that there was a different regime, namely the deflection of small molecules at low
temperatures, which should allow the spatial separation of quantum states~\cite{Stern:ZP39:751}.
Rabi developed this into the molecular beam resonance method~\cite{Rabi:PR55:526}. Later, multipole
focusers where used as quantum-state-specific lenses for molecules in so-called low-field-seeking
states~\cite{Gordon:PR95:282, Bennewitz:ZP141:6}, used in very-high-resolution
spectroscopy, for instance, of ammonia~\cite{Gordon:PR99:1253}, and to experimentally implement the
MASER~\cite{Gordon:PR99:1264}. Similar experiments with decelerated beams of ammonia increased the
achievable resolution further~\cite{Veldhoven:EPJD31:337}. Many of these methods were used to study
the properties of separated individual molecular species, if ``only'' because they created samples
of molecules in a single (internal) quantum state.

The rapid and extreme cooling provided by supersonic expansions of seeded molecular beams, down
below 1~K~\cite{Scoles:MolBeam:1and2, Hillenkamp:JCP118:8699}, resulting in only few quantum states
populated for polyatomic molecules, allowed to extend these techniques. More recently, over the last
fifteen years, the field of ultracold molecules and cold chemistry propelled molecular physics and
physical chemistry into new regimes, extensively described in a special issue on ``Ultracold
Molecules'' in Chemical Reviews~\cite{Jin:CR112:4801}. These reviews describe wonderful experiments
and avenues into ultracold physics and chemistry, including practically fully controlled
quantum-state specific reactions. However, the discussion is focused on small molecules and the
issues of molecular species, such as structural isomers, is left out. This is likely due to the fact
that the control of complex molecules is less advanced.

The manipulation of complex molecules with electric fields is hampered by the fact that their
quantum states are all strong-field seeking under practically useful conditions. This is not of
severe influence for deflection experiments, which this review focuses on. However, it makes
focusing~\cite{Auerbach:JCP45:2160, Guenther:ZPCNF80:155, Filsinger:PRL100:133003,
   Filsinger:PRA82:052513, Putzke:PCCP13:18962} and deceleration~\cite{Bethlem:PRL88:133003,
   Tarbutt:PRL92:173002, Bethlem:JPB39:R263, Wohlfart:PRA78:033421, Wohlfart:PRA77:031404}
cumbersome, at least. Previous overviews of such experiments are given
elsewhere~\cite{Kuepper:FD142:155, Schnell:ACIE48:6010, Filsinger:PCCP13:2076,
   Meerakker:CR112:2012}. Alternatively, microwave~\cite{Odashima:PRL104:253001} and
optical~\cite{Stapelfeldt:PRL79:2787, Zhao:PRL85:2705, Fulton:NatPhys2:465} ac fields were used to
manipulate the motion of neutral molecules. Similarly, connected by the seminal work of Arthur
Ashkin~\cite{Ashkin:PRL24:156}, very large, tens-of-nanometer- to micrometer-size ``molecules'' can
be manipulated using radiation pressure or photophoresis \emph{in
   vacuo}~\cite{Lebedew:AnaPhys311:433, Ehrenhaft:AnaPhys361:81, Eckerskorn:OE21:30492,
   COMOTION:website}.

\subsection{Detecting molecular species}
Starting from the gas-phase-spectroscopic characterization of a second conformer of
glycine~\cite{Suenram:JACS102:7180} and the observation of multiple conformers of tryptophan even in
cold supersonic expansions~\cite{Rizzo:JCP84:2534}, detailed spectroscopic characterizations of
these molecular species have been performed~\cite[and references therein]{Vries:ARPC58:585,
   Neill:PCCP13:7253}. Many of these studies provide partial structural information or allow to
benchmark quantum-chemistry calculations, which in turn provide structures of individual species.
Moreover, these high-resolution spectroscopies allow to distinguish different species by their
distinct resonance frequencies.

Alternative approaches to the investigation of the structure and dynamics of molecules are
diffractive imaging techniques. Traditionally, x-ray crystallography and gas-phase electron
diffraction (GED) have successfully derived structures, including those of
DNA~\cite{Watson:Nature171:737} and fullerene~\cite{Hedberg:Science254:410}, respectively.
Time-resolved approaches promise to allow the recording of so-called ``molecular movies'' of
molecules \emph{in action}~\cite{Williamson:Nature386:159, Sciaini:RPP74:096101,
   Neutze:Nature406:752, Barty:ARPC64:415}. However, these techniques are not molecular species
selective and, therefore, purified samples need to be provided~\cite{Filsinger:PCCP13:2076,
   Barty:ARPC64:415}. Moreover, in order to increase the obtainable information, samples of
molecules fixed in space are highly beneficial and proof-of-principle experiments have been
performed~\cite{Hensley:PRL109:133202, Kuepper:PRL112:083002}.

Molecular-frame photoelectron angular distributions (MFPADs)~\cite{Reid:MolPhys3:131} provide an
alternative approach to image electronic structure. Direct photoelectrons from strong-field
ionization image the electron density of the highest occupied molecular orbitals, including clear
pictures of their symmetries~\cite{Bisgaard:Science323:1464, Holmegaard:NatPhys6:428,
   Akagi:Science325:1364, Hansen:PRL106:073001, Maurer:PRL109:123001}. Interference due to
rescattering electrons results in more intricate photoelectron angular distributions, which can be
considered as laser-induced electron diffraction when the wavelength of the returning electron is
short~\cite{Zuo:CPL159:313, Blaga:Nature483:194}. Alternatively, the photons generated by the
coherent recombination of the accelerated electron with the ion core can be analyzed in
high-harmonic-generation spectroscopy~\cite{Itatani:Nature432:867, Woerner:Nature466:604}.
Single-photon ionization can also provide a holographic electron diffraction approach to the
measurements of molecular structure~\cite{Landers:PRL87:013002}, which is applicable to complex
molecules when they are strongly aligned or oriented~\cite{Krasniqi:PRA81:033411,
   Boll:PRA88:061402}.

\subsection{Chemistry of  molecular species}
The observation, understanding, and, eventually, control of quantum-state specificity of chemical
reactions has been the holy grail of chemistry for a century, and it has repeatedly been approached
by different communities. Important milestones were the state-to-state reactive scattering
experiments starting about fifty years ago~\cite{Lee:Science236:4803}, the investigation of
so-called half-collisions of molecular clusters~\cite{Zewail:JPCA104:5660}, and the
shaped-laser-pulses coherent-control approaches~\cite{Warren:Science259:1581,
   Assion:Science282:919}. Successful experiments were always limited to small molecular systems,
largely due to the failure to control large molecules well enough. This is combined with the
difficulties to appropriately theoretically describe or model the chemistry, which is even worse for
poorly defined initial conditions. Nevertheless, extremely detailed insight into chemical reactions
was obtained from state-specific, cold collision studies~\cite{Wang:Science342:1499,
   Vogels:PRL113:263202} as well as from ultrafast imaging experiments~\cite{Ihee:Science291:458,
   Miller:Science343:1108}.

In the remainder of this manuscript, we describe an approach to control molecules that opens new
avenues for studies of the intrinsic molecular structure and dynamics of complex molecules, with the
ultimate goal to develop a detailed and deep understanding of molecular function and the
structure-function relationship. We give an account of the electric deflection method for the
spatial separation of molecular species and quantum states and we outline the applicability of the
produced controlled samples to further our understanding of these individual chemical species. In
\autoref{sec:possibilities} we describe the basic concepts of dc-electric-field based techniques for
the manipulation of the motion and selection of individual species of neutral molecules. In
\autoref{sec:applications}, we highlight selected applications in physical chemistry, and in
\autoref{sec:outlook} we propose further experiments that are enabled by the current technology.

\section{Background}%
\label{sec:possibilities}%
By exploiting the Stark effect, strong electric fields provide a handle on neutral molecules,
allowing for the manipulation of their motion. Here, we describe the physics of the interaction of
neutral molecules with dc electric fields. Since this interaction is quantum-state specific, it
allows, for instance, to spatially separate molecular beams according to quantum state.
Appropriately applied, this results in the spatial separation of molecular species.

Generally these techniques are applicable to all molecules, due to the invariably non-zero
polarizabilities, but for non-polar molecules the interaction strength with electric fields is
typically small. Therefore, we restrict our discussion to polar molecules and to the
electric-dipole-moment interaction with the electric field. Moreover, while we point out alternative
techniques, we will focus our discussion on electric-field techniques and the simplest electric
manipulation device, the electric deflector.

\subsection{Stark effect primer}
\label{sec:poss:stark-effect}
\begin{figure}
   \centering
   \includegraphics[width=0.25\linewidth]{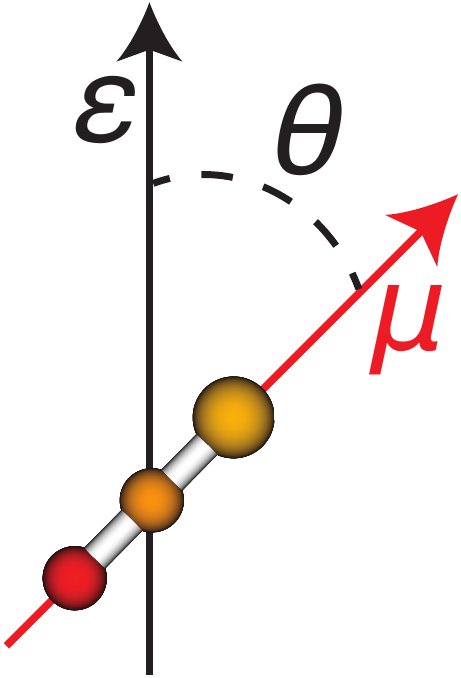}
   \caption{Electric field $\vec{\epsilon}$, electric dipole moment $\vec{\mu}$, and angle
      $\theta$ between them, depicted for the OCS molecule.}
   \label{fig:vector-sketch}
\end{figure}

A polar molecule has a so-called permanent electric dipole moment, \ie, the centers of positive and
negative electric charges are separated, in a well-defined way, in the molecular frame. This
dipole's interaction with an external electric field, which results in energy shifts and
wavefunction hybridization, is called the Stark effect. The resulting energy shift is given by
\begin{equation}
   W = -\vec{\mu}\cdot\vec{\epsilon} = -\mu\,\epsilon\expectation{\cos\theta} = -\mueff\,\epsilon
   \label{eq:stark-energy}
\end{equation}
with the dipole moment $\vec{\mu}$ (absolute value $\mu$), the electric field
$\vec{\epsilon}$ (absolute value $\epsilon$), and the angle $\theta$ between the two directions, as
shown in \autoref{fig:vector-sketch}.
The angular probability distribution and, therefore, the expectation value
$\expectation{\cos\theta}$ is defined by the molecule's rotational
wavefunction~\cite{Gordy:MWMolSpec}. The resulting effective dipole moment along the electric field
axis is the space-fixed dipole moment $\mueff=\mu\expectation{\cos\theta}$ of the molecule. Since
$\expectation{\cos\theta}$ is a property of the quantum state, $\mueff$ is as well.

For a molecule in field-free space the different projections $M$ of the angular momentum onto the
laboratory axis are degenerate. The angular probability density is spherical, \ie,
$\expectation{\cos\theta}=0$ and the molecule's space-fixed dipole moment vanishes. The Stark effect
lifts this degeneracy and mixes the wavefunctions, resulting in $\expectation{\cos\theta}\neq0$.
Furthermore, the energy of the system depends on the electric field strength $\epsilon$. Therefore,
the molecule will experience a force in an inhomogeneous electric field: depending on the sign of
$\expectation{\cos\theta}$ it will be attracted towards regions of weaker or stronger electric
field. Thus, providing appropriately shaped inhomogeneous electric fields enables the manipulation
of the motion of molecules in space.

Theoretically, we separate the electronic and nuclear degrees of freedom through the
Born-Oppenheimer approximation and the Eckart conditions~\cite{Gordy:MWMolSpec}. For rigid
closed-shell molecules in moderately strong electric fields, we can then describe the Stark
interaction solely in the rotational degrees of freedom of the molecule. The rotational
wavefunctions couple the molecular and laboratory frames, \eg, the (molecular-frame-fixed)
dipole moment $\vec{\mu}$ and the (laboratory-frame-fixed) effective dipole moment $\mueff$. For a
polar molecule, this is described by an Hamiltonian $\op{H}$
\begin{equation}
   \op{H} = \op{H}_{\mathrm{rot}} + \op{H}_{\mathrm{Stark}}
   \label{eq:hamiltonian}
\end{equation}
where $\op{H}_{\text{rot}}$ is the field-free rotor Hamiltonian and $\op{H}_{\text{Stark}}$
represents the Stark interaction~\cite{Gordy:MWMolSpec, Chang:CPC185:339}:
\begin{equation}
   \op{H}_{\text{Stark}}=-\vec{\mu}\cdot\vec{\epsilon} = -\epsilon\sum_{g=x,y,z}\phi_{Zg}\,\mu_g
\label{eq:Hstark}
\end{equation}
where $x,y,z$ represent a molecule-fixed coordinate system, $\mu_g$ represents the dipole moment
components along the molecule-fixed axes $x,y,z$, and $\phi_{Zg}$ are the direction cosines of the
molecular axes with reference to the space-fixed $X,Y,Z$-axes, with $Z$ being oriented along the
electric-field direction. Corresponding matrix elements to numerically set up the Hamiltonian matrix
are given elsewhere~\cite{Chang:CPC185:339}. Higher-order effects, \eg, the polarizability, can
mostly be neglected in the case of polar molecules, although they can become relevant in very strong
fields. For instance, the dipole moment
($\mu=4.5152~\text{D}$~\cite{Wohlfart:JMolSpec247:119}) of benzonitrile's ground state leads to an
energy shift of $\sim\!300$~GHz at 200~kV/cm, but the corresponding effect due to the polarizability
of the, very similar, non-polar molecule benzene is only 50~MHz~\cite{Okruss:JCP110:10393}, \ie,
almost four orders of magnitude smaller.

\begin{figure}[t]
   \centering
   \subfigure[OCS]{\includegraphics[width=0.47\linewidth]{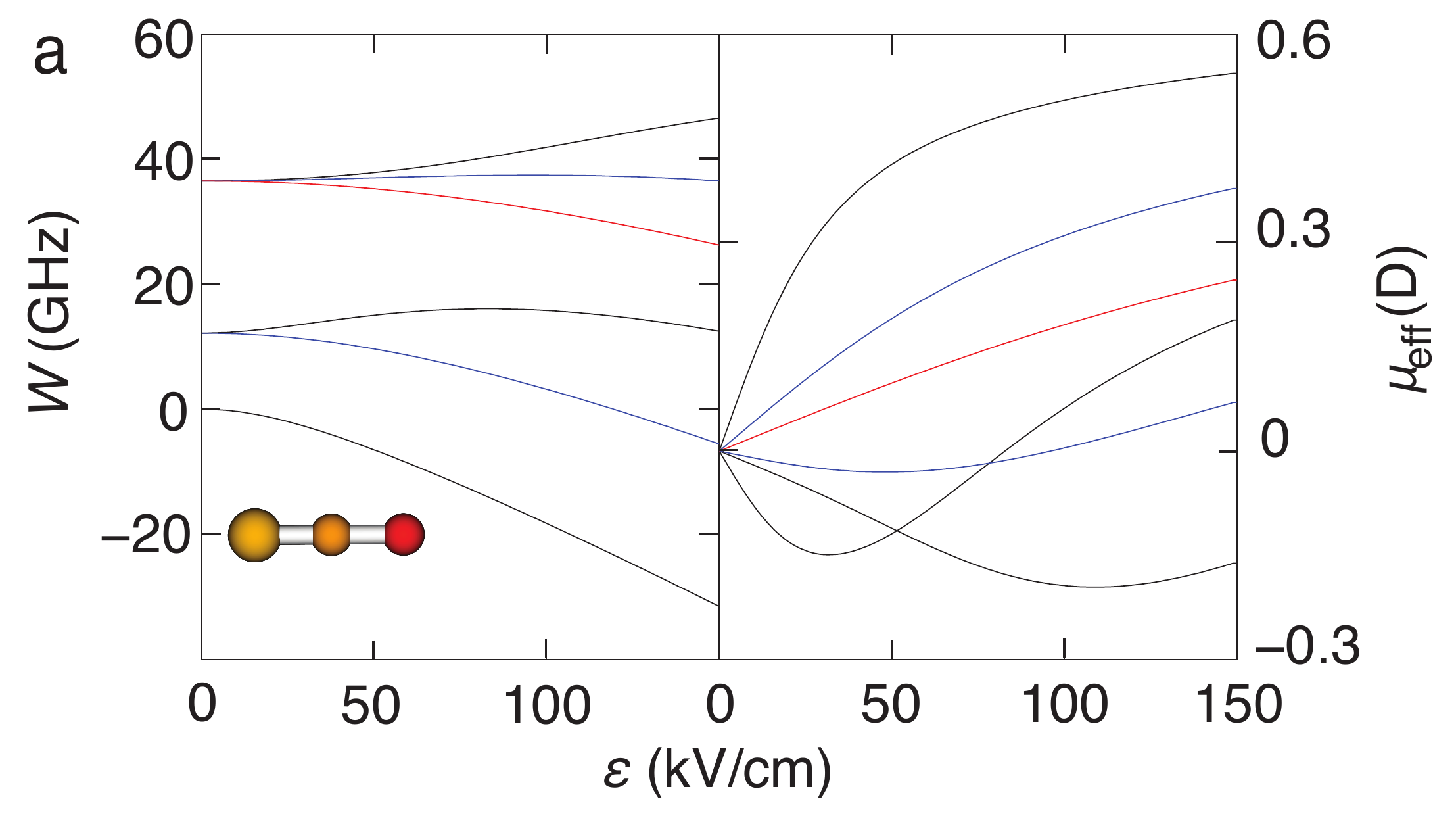}\label{fig:sel:OCS_Stark_curves}}
   \hfill%
   \subfigure[indole]{\includegraphics[width=0.47\linewidth]{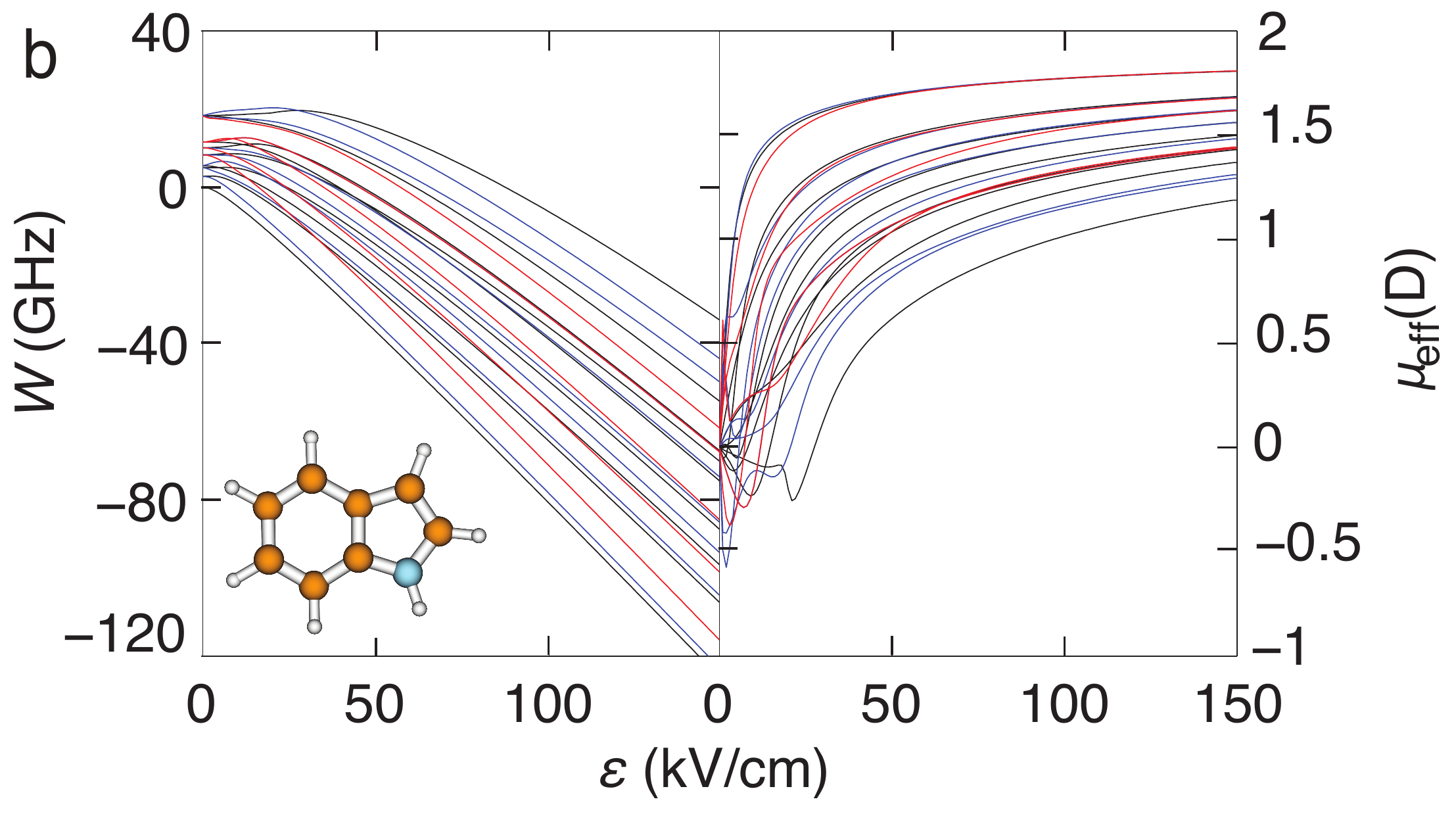}\label{fig:sel:Ind_Stark_curves}}
   \caption{Stark energies $W$ (left plots) and effective dipole moments $\mueff$ (right plots) of
      (a)~the linear-top molecule OCS and (b)~the asymmetric-top molecule indole (C$_8$NH$_7$) for
      the $M=0$ (black), $M=1$ (blue), and $M=2$ (red) levels of all $\tilde{J}=0$--$2$ states. All
      data were calculated using CMIstark~\cite{Chang:CPC185:339}. See text for details.}
   \label{fig:sel:Stark_curves}
\end{figure}

The energy levels of the molecule in the field, \ie, the Stark energies, are obtained by setting up
the Hamiltonian matrix in the Wang-symmetrized symmetric-top basis and subsequent numerical matrix
diagonalization for a number of electric field strengths. If the Hamiltonian matrix is appropriately
block diagonalized exploiting the full symmetry of the problem, adiabatic Stark energy curves can
simply be derived by interpolation between the appropriate calculated
energies~\cite{Chang:CPC185:339}. These calculations, including the appropriate symmetrization, are
described in detail in reference~\citealp{Chang:CPC185:339}, which also provides a user-friendly
program to perform such numerical calculations for typical molecules. More analytical descriptions
of the basic principles are given, for instance, in reference~\citealp{Bethlem:JPB39:R263}. The
resulting Stark energy curves, depicting the quantum-state energies as a function of electric field
strengths, are shown in \autoref{fig:sel:Stark_curves} for the linear molecule OCS and the
asymmetric-top molecule indole.
While only $M$ remains a good quantum number in a static laboratory frame electric field, we generally add labels
$\tilde{J}$ or $\tilde{J}_{\tilde{K}_a,\tilde{K}_c}$ for the linear top OCS and the asymmetric top
indole, respectively, to depict the adiabatically corresponding field-free states in order to
uniquely identify the molecular quantum states, this is detailed in
reference~\citealp{Chang:CPC185:339}. Moreover, we also plot the effective dipole moment functions
for the same quantum states in \autoref{fig:sel:Stark_curves}, which are defined as
\begin{equation}
   \mueff(\epsilon) = -\frac{\partial{W}}{\partial{\epsilon}}
   \label{eq:mueff-general}
\end{equation}

From \autoref{fig:sel:Stark_curves}\,a it is obvious that the Stark energies $W$ can decrease or
increase as a function of electric field strength $\epsilon$ and, correspondingly, the effective
dipole moments $\mueff(\epsilon)$ can be positive or negative. Thus, depending on the quantum state,
the molecules orient or anti-orient their dipole moments with respect to the electric field and
molecules in these states are attracted to regions of stronger or weaker electric field,
respectively. We call these states strong-field-seeking and weak-field-seeking states, which are
traditionally abbreviated as ``hfs'' and ``lfs'', respectively.\footnote{These abbreviations are
   used due to the originally, traditionally, used terms ``high-field seeking'' (hfs) and
   ``low-field seeking'' (lfs) quantum states.}

As the field strength increases, so does the number of coupled $J$ states with a significant
contribution to the formed pendular states, eventually turning all low-energy states into hfs
states, \ie, they start to behave like classical dipoles. Generally, the field strength at which this
occurs is smaller for larger molecules, with larger moments of inertia (smaller rotational
constants) and for higher-energy states, but it depends on the exact energies and coupling
strengths. Moreover, avoided crossings can lead to quasi-chaotic behavior and have to be taken into
account~\cite{AbdElRahim:JPCA109:8507, Chang:CPC185:339}.

The plots for OCS and indole in \autoref{fig:sel:Stark_curves} show qualitatively distinct behaviour
for the plotted, and practically achievable, range of electric field strengths, highlighting the
differences mentioned above. For OCS all rotational states are well separated, although the
$\ket{JM}=\ket{1,0}$ state shows a turnaround from lfs to hfs state at about 83~kV/cm, which is due
to the initially dominant coupling to the \ket{0,0} state getting overwhelmed by the coupling to the
\ket{2,0} (and higher) states. Nevertheless, for these experimentally achievable field strengths the
typically populated states partly show hfs and partly lfs behaviour. To the contrary, for indole
these effects all occur at much smaller field strengths and under practically relevant field
strengths, 50--150~kV/cm, all states are hfs. This is due to the significantly smaller rotational
constants and the much larger number of states ``per $J$'' of the asymmetric top compared to the
linear top OCS. Together, these effects result in a much larger density of states, correspondingly
stronger Stark mixing, and, thus, the fact that all states become hfs in relatively weak fields,
\ie, $\epsilon<25$~kV/cm, already.

\subsection{Molecules in inhomogeneous electric fields}
\label{sec:molecules-in-fields}
The field-strength dependence of the Stark energy, coupled with the principle of minimum energy,
allows the manipulation of the motion of molecules using appropriately shaped inhomogeneous electric
fields. The force $\vec{F}$ exerted on the molecule is
\begin{equation}
   \vec{F} = -\vec{\nabla}W = \mueff(\epsilon)\cdot\vec{\nabla}\epsilon
   \label{eq:deflection_force}
\end{equation}
This force is exploited in the electric deflector to disperse a molecular beam according to the
molecules' effective dipole moments \mueff, \ie, according to quantum state. An ideal electric
deflection field would exert a strong and constant force in one direction, while the force in the
perpendicular direction would be zero. According to Maxwell's equations, this is not possible for
hfs molecules, but one can practically approximate this case using so-called two-wire
fields~\cite{Ramsey:MolBeam:1956}.

States with different \mueff experience different deflection forces in the field, thus, traveling
through the field they acquire different transverse velocities according to their dipole moment to
mass ratio. Furthermore, since structural isomers of complex molecules and even molecular clusters
of different sizes can have distinct dipole moments, and hence dipole moment to mass ratios, these
species are transversely separated. More specifically, the species with the largest dipole moment to
mass ratio is deflected the most and can cleanly be separated from all other components of the
molecular beam~\cite{Filsinger:JCP131:064309}. This can be generalized to the most polar structural
isomer of a complex molecule, which is deflected the most of all existing conformers and can easily
be purified (see \autoref{sec:application:conformer-selection}).

To create inhomogeneous electric fields that provide the best forces for the manipulation of the
transverse motion, a number of electric field geometries have been devised. An analysis of the
electrostatic potential $\Phi$ and deflection and focusing fields $\vec{\epsilon}=-\vec{\nabla}\Phi$
can be performed in terms of a multipole expansion~\cite{Kalnins:RSI73:2557, Bethlem:JPB39:R263,
   DeNijs:PCCP13:19052}. Using that formalism in two dimensions ($X,Y$), $\Phi$ is expressed
as~\cite{DeNijs:PCCP13:19052}:%
\begin{equation} 
   \Phi(X,Y) = \Phi_0\left[\sum_{n=1}^{\infty}\frac{a_n}{n}\left(\frac{r}{r_0}\right)^n\sin(n\theta)+
      \sum_{n=1}^{\infty}\frac{b_n}{n}\left(\frac{r}{r_0}\right)^n\cos(n\theta)\right]
\end{equation}
using the usual cylindrical coordinates $r=\sqrt{(X^2+Y^2)}$ and $\theta=\arctan(Y/X)$. $a_n$ and
$b_n$ are the multipole expansion coefficients and $r_0$ and $\theta_0$ are scaling factors that
characterize the size of the electrode structure and the applied voltages, respectively.

\begin{figure}[t]
   \centering
   \includegraphics[width=0.8\linewidth]{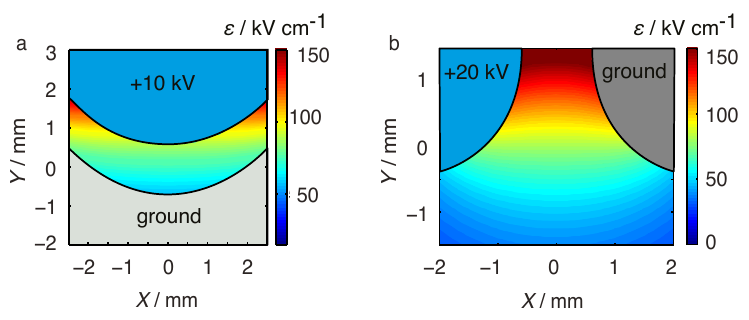}
   \caption{Two possible deflector electrode geometries: a) conventional (traditional) and b) wedge
      high-voltage-electrode arrangements and resulting deflection fields. The colour map indicates
      the magnitude of the electric field in the gap between deflector electrodes. The blue and grey
      areas represents the shapes of deflector electrodes, \ie, the geometric boundaries of the
      deflection field. The geometry on the left corresponds to the deflector and field depicted in
      \autoref{fig:applications:reactions} and it is the geometry used in most experimental
      setups~\cite{Ramsey:MolBeam:1956, Miller:AAMOP25:37, Schaefer:PRL76:471,
         deHeer:arXiv0901:4810, Filsinger:JCP131:064309, Trippel:MP111:1738,
         deNijs:JMolSpec300:79}.}
   \label{fig:sel:deflector:geometry}
\end{figure}
\autoref{fig:sel:deflector:geometry} shows two typical deflection fields, which are created by
setting either all $b_n=0$ and \mbox{$|a_1|\gg|a_2|\gg|a_3|>0$},~\mbox{$a_n=0~\forall~n>3$} or,
alternatively, all $a_n=0$ and \mbox{$|b_1|\gg|b_2|\gg|b_3|>0$},~\mbox{$b_n=0~\forall~n>3$}. Both
configurations correspond to two-wire fields~\cite{Ramsey:MolBeam:1956}. They exhibit large
gradients along one axis and are nearly homogeneous along the perpendicular axis. Although the
resulting potentials and electric fields for the two deflector geometries are different, the
magnitude of the electric field and forces at the region suitable for the molecular beam deflection
are the same.

Following reference~\citealp{DeNijs:PCCP13:19052}, the electrostatic potential and the resulting
electric field norm for the field in \autoref{fig:sel:deflector:geometry}\,(a) can be written as
\begin{align} 
   \Phi (X,Y) & = \Phi_0 \left(a_1 \frac{Y}{r_0} + a_2 \frac{Y^2-X^2}{2r^2_0} + a_3
      \frac{Y^3-3YX^2}{3r^3_0}\right)
   \label{eq:AA_potential} \\
   \epsilon(X,Y) &= \sqrt{\left(\frac{\partial\Phi}{\partial X}\right)^2+\left(\frac{\partial\Phi}{\partial Y}\right)^2}
\end{align}
Throughout the region $r<r_0$, we can expand the force resulting from \eqref{eq:AA_potential} as
\begin{align} 
  F_X(X,Y) &= \mueff \epsilon_0 \left[\left(\left(\frac{a_2}{a_1}\right)^2 - 2\frac{a_3}{a_1}\right)\frac{X}{r^2_0} -
  \left(\left(\frac{a_2}{a_1}\right)^3 - 4\frac{a_2}{a_1}\frac{a_3}{a_1}\right)\frac{XY}{r^3_0} + \cdot\cdot\cdot \right]
  \label{eq:Fx_stark} \\
  F_Y(X,Y) &= \mueff \epsilon_0 \left[\frac{a_2}{a_1}\frac{1}{r_0} + 2\frac{a_3}{a_1}\frac{Y}{r_0^2} - \left(
  \frac{1}{2}\left(\frac{a_2}{a_1}\right)^3 - 2\frac{a_2}{a_1}\frac{a_3}{a_1}\right)\frac{X^2}{r^3_0} + \cdot\cdot\cdot \right]
   \label{eq:Fy_stark}
\end{align}
where $\epsilon_0 = (\Phi_0/r_0)\sqrt{a_1^2+b_1^2}$. To obtain fields that are close to the ideal
case described above, coefficients should be chosen such that the first term in \eqref{eq:Fy_stark}
is the only significantly contributing term, while all higher-order terms, and terms in
\eqref{eq:Fx_stark} should vanish. The first term in \eqref{eq:Fy_stark}, the desired term, scales
as $a_2/a_1$ while the other undesired terms scale as $a_3/a_1$ or as the second or third power of
$a_2/a_1$. We can make these undesired terms arbitrary small by choosing $a_3=0$ and $a_2/a_1\ll1$,
but only at the expense of the strength of the deflection force. However, in practice one cannot
afford to choose $a_2/a_1$ much smaller than $1/5$~\cite{DeNijs:PCCP13:19052}. The dominant
undesired term in this case is the first term of \autoref{eq:Fx_stark}. This term can be canceled
with an appropriate choice of $a_3$, but this introduces other unwanted terms. Overall, for
molecules in hfs states, deflection fields can provide focusing in one transverse direction, but not
in both directions~\cite{Bethlem:JPB39:R263}. For example, for the deflector geometry shown in
\autoref{fig:sel:deflector:geometry}\,(a), where the value of $a_3$ is positive, molecules in hfs
states are focused in the $X$ direction, but they are slightly defocused in the $Y$
direction~\cite{DeNijs:PCCP13:19052}. When the value of $a_3$ is negative, such as the deflector
geometry shown in \autoref{fig:sel:deflector:geometry}\,(b), molecules in hfs states are focused
(defocused) along the $Y$-axis ($X$-axis) instead. Typical deflectors used in actual experimental
setups~\cite{Ramsey:MolBeam:1956, Filsinger:JCP131:064309, Trippel:MP111:1738,
   deNijs:JMolSpec300:79} have geometries that correspond approximately to $a_1=0.5$, $a_2=0.49$,
$a_3=0.42$. However, the analytical model does not describe the electric field well away from the
axis ($r\ne0$), and in practice numerically generated electric fields are used to model the
experiments~\cite{Filsinger:JCP131:064309}.

While focusing of hfs molecules cannot be achieved using static electric fields,
alternating-gradient, so-called strong focusing, can be applied for the dynamic focusing of neutral
molecules in hfs states~\cite{Auerbach:JCP45:2160, Bethlem:JPB39:R263, Filsinger:PRL100:133003}.
Advanced concepts allow even for the deceleration of these molecules~\cite{Auerbach:JCP45:2160,
   Bethlem:PRL88:133003, Tarbutt:PRL92:173002, Wohlfart:PRA77:031404, Wohlfart:PRA78:033421}. While
these techniques are directly relevant for the topic discussed in this review, a detailed account is
beyond the scope of this article. Extensive treatments can be found in the
literature~\cite{Auerbach:JCP45:2160, Bethlem:JPB39:R263, Filsinger:PRA82:052513} and their use as
strong-focusing storage rings has been proposed~\cite{Nishimura:RSI74:3271, DeNijs:PCCP13:19052}.
Since such a storage ring could, possibly simultaneously, confine molecules in hfs and lfs states,
they would offer interesting options for collision studies. However, these techniques for molecules
in hfs states have not been explored much in actual applications, most likely due to their
significant experimental complexity~\cite{Bethlem:JPB39:R263, Wohlfart:PRA78:033421,
   Putzke:PCCP13:18962, Putzke:thesis:2012}. Nevertheless, in the following discussions of the
applications of controlled molecules, we will point out results from strong-focusing manipulation
where available.

\subsection{State selection by deflection}
\label{sec:theory-state-selection}
Following this discussion, we can exploit the quantum-state-specific Stark effect of molecules in
inhomogeneous electric fields to spatially separate molecules according to quantum state. For a
given, \eg, static, electric field with an approximately spatially constant electric field gradient,
such as the two-wire fields described in \autoref{sec:molecules-in-fields}, the force exerted on a
molecule is proportional to the effective dipole moment (at the given field strength). Therefore,
the deflection depends monotonically on the effective dipole moment. For a beam of molecules, which
enter the deflection field all with similar speeds, this deflection field acts as a prism that
disperses the molecules in the beam according to \mueff, \ie, according to quantum state. Already
proposed by Otto Stern in 1926~\cite{Stern:ZP39:751} for beams of small molecules at low
temperatures, this technique allows the spatial separation of individual quantum states. It is also
directly applicable to larger molecules, as shown here, as it always disperses the molecular beam
accordingly. However, in order to create useful isolated samples, with a small subset of specific
quantum states, it is of utmost importance to work with initially cold beams. These can be created,
with temperatures in the moving frame reaching below 1~K, in high-pressure supersonic
expansions~\cite{Scoles:MolBeam:1and2, Hillenkamp:JCP118:8699}. It avoids issues with non-rigidity,
\eg, internal motions of the molecules~\cite{Trippel:PRA86:033202, AbdElRahim:JPCA109:8507}.
Correspondingly, different structural isomers are frozen and separable, as described in
\autoref{sec:application:conformer-selection}. Moreover, the lowest-energy rotational states are the
states with the smallest angular momentum, making them most suitable for orientation experiments,
\ie, they have the largest effective dipole moments \mueff, and thus the strongest Stark
interaction. Therefore, they show the strongest response to the electric field. Since the deflection
method is purely selecting molecules from the original distribution, the original number of
molecules in these states, which is increased by lowering the temperature, determines the number of
molecules available for an experiment with state-selected samples.

In the following, we show the results of Monte Carlo trajectory simulations of OCS molecules in a
molecular beam to demonstrate state selection using electrostatic deflection fields. We also provide
the source code of a prototype trajectory simulation program, with more details given in the
supplementary information. In these simulations, we employ the deflection field geometry shown in
\autoref{fig:sel:deflector:geometry}\,(a), with a 15\,cm long deflector, a gap between the deflector
electrodes of 1.4~mm, and a radius of the rod and trough of 3.0~mm and 3.2~mm,
respectively~\cite{Filsinger:JCP131:064309}. An aperture, typically a conical skimmer, is placed
about 3~cm before the deflector with an orifice diameter smaller than the gap between the
electrodes, \eg, 1~mm. The simulations take into account these geometric boundaries of the setup.
When the molecular beam is parallel to the deflector, the mean velocities along the transverse, $X$
and $Y$, axes are zero and the corresponding velocity spreads of a supersonically expanded skimmed
beam are typically within a few m/s~\cite{Luria:JPCA115:7362}. The pulsed valve is typically located
some ten centimeter before the deflector. The acceleration force acting on a molecule at position
$(X,Y,Z)$ is calculated as the product of the gradient of the deflector electric field
$\vec{\nabla}\epsilon(X,Y,Z)$ and the effective dipole moment $\mueff(\epsilon)$, which is a
function of the electric field strength $\epsilon(X,Y,Z)$, see \eqref{eq:deflection_force}. For
typical experimental conditions and a molecular dipole moment of a few Debye, this results in a
transverse velocity component of a few m/s. This results in a deflection amplitude on the order of
1~mm after a successive free-flight distance of 20~cm. The full deflection profiles, \ie, the
density distribution at some position along the beam, are calculated for individual quantum states
and these profiles are added according to their populations in the incoming beam, including the
Boltzmann distribution, degeneracies, and nuclear-spin-statistical weights.

\begin{figure}[t]
   \centering
   \includegraphics[width=0.5\linewidth]{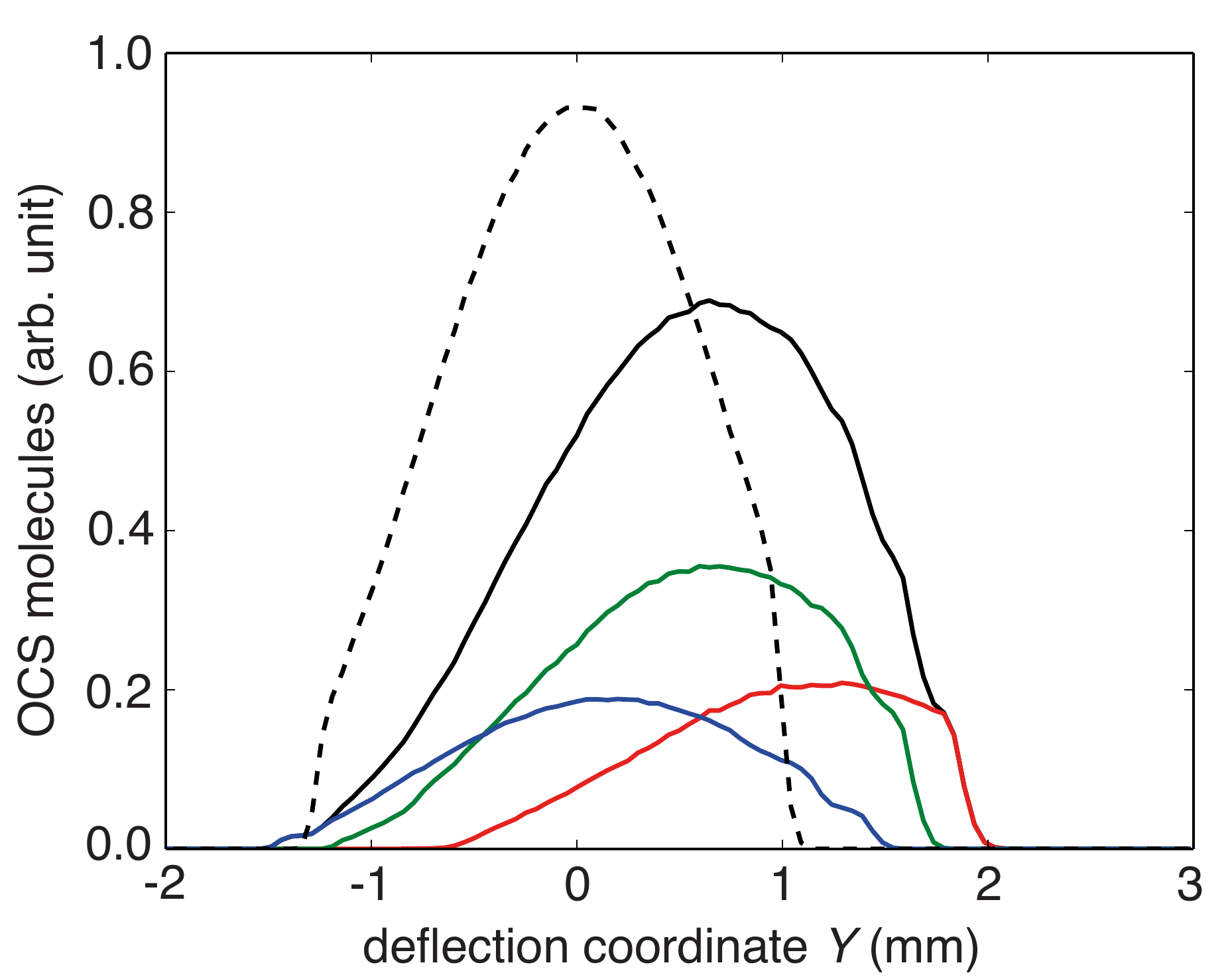}
   \caption{OCS deflection simulations of undeflected beam profile (black dash line) and with a
      deflection voltage of 20\,kV (black solid line) at 1\,K. Single state deflection profiles:
      $J=0$ (red), $J=1$ (green) and $J=2$ (blue). See text for more details.}
   \label{fig:sel:OCS_def}
\end{figure}
\autoref{fig:sel:OCS_def} shows the simulated deflection profiles for a beam of OCS with an initial
speed of 1900~m/s, corresponding to a seeded helium expansion. The dashed black line depicts the
overall profile for the undeflected beam, without any field applied, which has an identical shape
for all states. Here, the gap between the deflector electrodes defines the shape and the width of
the undeflected beam profile. The coloured lines are the deflection profiles of the individual
rotational states $J=0,1,2$, summing up all $M$, for an applied deflection voltage of 20~kV. The
solid black line gives the weighted sum of these states for a rotational temperature of 1\,K. The
deflection profiles of individual rotational quantum states exhibit different shapes and deflection
amplitudes due to the distinct \mueff of their $M$ levels. Particularly, the rotational ground state
\ket{0,0} has the largest \mueff and, therefore, experiences the largest spatial deflection. These
experimental conditions allow the production of single-quantum-state samples, \eg, around the
deflection coordinate $Y=1.8$~mm, only the $J=0$ state is present in the deflected molecular beam.
Similarly, almost pure $J=2$ state samples can be obtained below $Y=-1.2$~mm and the deflection
downwards, to lower $Y$, is reminiscent of the lfs nature of the \ket{2,0} state at the field
strengths in the deflector. Better separation could be achieved using a smaller skimmer before the
deflector to provide narrower distributions for individual states. An experimental demonstration of
the production of individual quantum states is described in \autoref{sec:quantumstates}.

For the propagation of molecules in inhomogeneous (or time-varying) electric fields one has to
carefully consider the adiabaticity of the process~\cite{Escribano:PRA62:023407, Wall:PRA81:033414},
\ie, the assumption that molecules stay in a single well-defined adiabatic quantum state. When the
rate of change of the electric field, amplitude or direction, become too fast, non-adiabatic
transitions will occur and alter the populations and the deflection profiles. Generally, when
starting with a cold sample this will result in a warming of the beam, on average reducing the
deflection amplitude~\cite{AbdElRahim:JPCA109:8507}. Moreover, the complicated and dense hyperfine
structure underlying the rotational-states results in complicated ``partial adiabaticities'', which
is due to the fact that different hyperfine levels of two rotational states might exhibit adiabatic,
partially adiabatic (mesobatic), or diabatic behaviour~\cite{Kirste:PRA79:051401,
   Veldhoven:PRA66:032501}. In the case of OCS, discussed here, this is not relevant as it does not
have nuclear spin and thus no hyperfine splittings, and in the relevant electric field strengths the
rotational states are well separated. Therefore, these molecules traverse through the experiment
adiabatically. For prototypical large molecules, such as indole and 3-aminophenol (see
\autoref{sec:application:conformer-selection}), these effects seem to be small enough to be
negligible for our electric deflection experiments. In \autoref{sec:outlook:nonadiabatic-effects} we
propose some experiments to further investigate this behaviour.

We point out that the deflector, discussed here, is the simplest device in a whole series. More
control can be achieved using the alternating-gradient focuser~\cite{Filsinger:PRL100:133003} and
decelerator~\cite{Auerbach:JCP45:2160, Bethlem:PRL88:133003, Bethlem:JPB39:R263} -- yielding the
complete line of devices equivalent to the bender, focuser, and LINAC for charged particles,
respectively. While some applications of the more complicated -- theoretically and experimentally --
dynamic-focusing devices are discussed in \autoref{sec:applications}, so far the deflector has made
the largest impact in the manipulation of molecules in hfs states, mostly for its simplicity that
allows to combine it with advanced further techniques. Moreover, the deflector has the advantage
that it fully separates the manipulated molecules from the original beam. However, for higher
selectivity advanced techniques need to be employed, \cf \autoref{sec:outlook:cyclic}, but under
these conditions dynamic focusing devices will become competitive again. A detailed discussion of
these effects is beyond the scope of this paper.

The resolving power of the deflector is, as such, not well defined. Its resolution depends, for
instance, on the speed of the molecules in the beam, on the size and shape of the deflector, and on
the probed interaction volume, the latter being comparable to the slit of an optical spectrometer.
Moreover, the achievable purity is a direct trade off with the corresponding density. The purity can
also be adjusted by simply changing a skimmer-orifice size before or after the deflector. Under
realistic experimental conditions, such as in the experiments described in this review, the
resolution of the deflector is comparable to that of the 1-m-long dynamic focuser, for which a
detailed investigation of the $\mu/m$ resolution has been carried out~\cite{Filsinger:thesis:2010,
   Filsinger:PRA82:052513}.

We can estimate the experimentally achievable separation of species from the various studies carried
out with the electrostatic deflector, which are summarised in \autoref{sec:applications}. Most of
these experiments employed a 15~cm-long 1.5-mm-opening deflector, with typical field strengths on
the order of 100~kV/cm. The interaction takes place around 20~cm behind the deflector. We note that
at larger distances the achieved purity will be greater, albeit at the expense of density. The
separation of individual quantum states of triatomic molecules with dipole moments on the order of
1--2~D, such as OCS~\cite{Nielsen:PCCP13:18971, Trippel:PRA89:051401R} or
H$_2$O~\cite{Horke:ACIE53:11965}, is feasible for differences in \mueff on the order of 0.1~D, which
are then spatially separated by $0.1-0.5$~mm. This is fully sufficient to address individual states
with a focused laser beam. While it is even harder to provide general rules for the separation of
structural isomers or clusters, we can estimate from the results presented in \ref{sec:applications}
that for molecules in the $m=100$--$200$~u range, cooled to $\sim\!1$~K, a dipole difference of 1~D
is sufficient to spatially separate them by around 1~mm~\cite{Filsinger:ACIE48:6900,
   Kierspel:CPL591:130, Trippel:PRA86:033202}. In principle, this separation can be further
increased through the use of slower molecular beams, \eg, heavier backing gases, or by sacrificing
target density and moving further into the tail of the deflected molecular beam distribution.

The density of molecules in the deflected beam depends on the density and velocity spread of the
original molecular beam and the mechanical transmissions of skimmers and the deflector. The
transmission through the deflector is only limited by its geometric size and thus only depends on
the collimation and the size of the initial molecular beam, and the deflection amplitude of
molecules. For the experimental setups generally used in our work, the transmission of the deflector
itself is typically over 90~\%, due to the application of tightly collimated molecular beams.
However, when applying voltages, the spatial dispersion of molecules may significantly modify the
transmission. This effect can even be used to deplete certain species from the interaction region
altogether~\cite{Kierspel:CPL591:130, Horke:ACIE53:11965}. In practice, using helium or neon buffer
gas, and with a total distance of 80--120~cm between the molecular beam valve and the interaction
region of the experiment, we achieve number densities in the deflected beam of
$10^8$--$10^9$~cm$^{-3}$~\cite{Filsinger:ACIE48:6900, Chang:Science342:98, Horke:ACIE53:11965}.

Nevertheless, while quantitative predictions are \emph{a priori} very difficult, we provide all
tools to predict the details for any given candidate system: Our code for the calculation of Stark
energy curves of linear, symmetric, and asymmetric top molecules is freely
available~\cite{Chang:CPC185:339}, and a script to simulate trajectories of molecules through an
electrostatic deflector is attached to this manuscript.

\section{Applications}
\label{sec:applications}

\subsection{Conformer selection: Investigating the structure-function relationship}
\label{sec:application:conformer-selection}
In order to develop a detailed understanding of the structure-function relationship, quantitative
gas-phase experiments, such as measurements of absolute cross-sections or reaction rates, are
essential. This requires the production of pure samples of individual conformers, that is, the
ability to separate molecular species with distinct structures, \cf \autoref{sec:introduction}. For
charged systems this has been demonstrated using ion-mobility schemes~\cite{Helden:Science267:1483,
   Jarrold:PCCP9:1659, Papadopoulos:FD150:243, Kanu:JMS43:1} or spectroscopic methods such as
selective ionisation of neutral precursors~\cite{Park:Nature415:306, Kim:Science315:1561}. While the
former is not applicable to neutral molecules, several spectroscopic schemes have been used for the
production of conformerically pure (or enriched) samples of neutral molecules in the gas-phase, \eg,
stimulated-emission pumping~\cite{Dian:Science303:1169} or IR-induced population
transfer~\cite{Dian:Science296:2369}. However, these spectroscopic methods are not generally
applicable for the separation of arbitrary neutral conformers and, furthermore, do not spatially
separate these. Thus, they do not allow non-species-specific investigations, such as strong-field
ionisation or diffractive imaging. These shortcomings are addressed by the separation techniques
using strong inhomogeneous electric fields. The first demonstration of conformer separation by
strong fields used alternating gradient focusers, through which, for a given ac frequency, only
molecules with selected $m/$\mueff ratios have stable trajectories~\cite{Filsinger:PRL100:133003,
   Filsinger:PRA82:052513}. This approach is, however, technologically extremely challenging. The
switching of large voltages (tens of kV) at kHz repetition rates is demanding, and the necessary
(multiple meter) long quadrupole electrodes are extremely sensitive to mechanical
misalignment~\cite{Bethlem:JPB39:R263, Wohlfart:thesis:2008, Putzke:thesis:2012}. An experimentally
much simpler device is the electrostatic deflector, which can be designed to produce a simple
two-wire field, as outlined in \autoref{sec:molecules-in-fields}~\cite{Ramsey:MolBeam:1956}. It is
generally applicable to all neutral molecules, including hfs states and even very large systems,
and, similar to alternating-gradient systems, spatially separates species on the basis of $m/$\mueff
ratio, but provides no radial focusing.

The selection of different conformers in the deflector is based on their differing dipole moments
due to a different arrangement of the functional groups in space. This has been demonstrated for
several small aromatic systems exhibiting two conformers~\cite{Filsinger:ACIE48:6900,
   Kierspel:CPL591:130, Horke:JoVE:e51137}.
\begin{figure}[t]
   \centering
   \includegraphics[width=\linewidth]{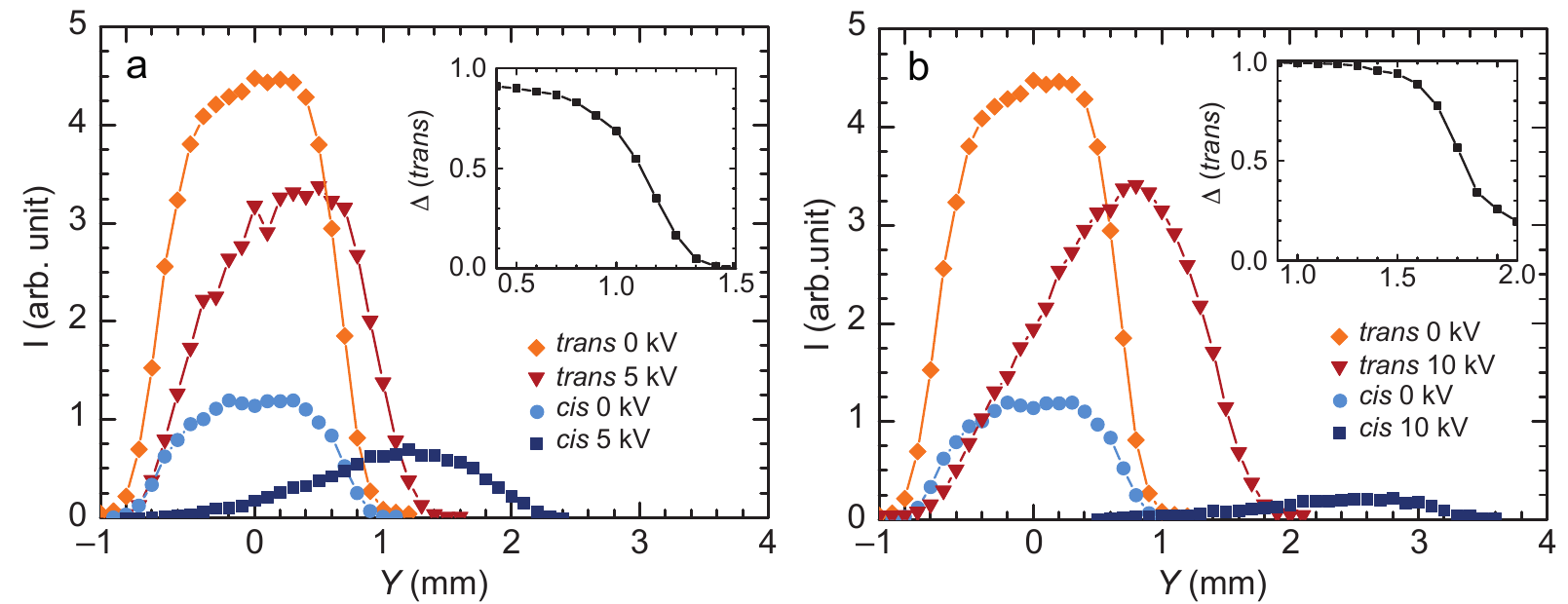}
   \caption{Separation of structural isomers of 3-aminophenol (3AP) with the electrostatic
      deflector. a) Due to its larger \mueff the \cis isomer experiences a larger deflection and
      pure samples can be obtained in a Neon beam at deflection voltages of 5~kV. b) Higher
      voltages, of 10~kV, can be used to remove all \cis population from the molecular beam, leaving
      a pure \trans sample. The insets show the fractional population $\Delta$ of \trans-3AP in the
      beams; see text for details.}
   \label{fig:applications:3AP}
\end{figure}
An example of the achieved separations and resulting conformeric purities is shown
in~\autoref{fig:applications:3AP} for the separation of \textit{cis}- and
\textit{trans}-3-aminophenol (3AP)~\cite{Filsinger:ACIE48:6900}. Here, pure samples of both
conformers can be produced and the species of interest selected \textit{via} the applied deflector
voltage. Using a Neon expansion, 5~kV is sufficient to separate the more polar \cis conformer from
the rest of the beam and obtain isolated samples, \autoref{fig:applications:3AP}\,a. The obtained
pure sample contains only the lowest energy quantum-states, as these experience the largest
deflection in the electric field, \cf \autoref{sec:poss:stark-effect}. In order to also produce pure
samples of the less polar conformer, in this case \trans 3AP, the voltage applied to the deflector
is increased further to 10~kV. At these voltages the \cis isomer is deflected so much the it does
not reach the interaction region anymore, \autoref{fig:applications:3AP}\,b. A sample containing
only the lowest rotational quantum-state of \trans is now produced in the region of $Y=1.4$~mm. A
similar conformer separation and production of rotationally cold samples has also been demonstrated
for 3-fluorophenol recently~\cite{Kierspel:CPL591:130}. This method is generally applicable to
systems with two populated conformers that differ in their dipole moment. The produced pure and cold
samples, which for smaller molecules can even be single quantum-states as shown in
\autoref{sec:quantumstates}, are beneficial in a variety of further experiments, such as molecular
alignment and orientation control~\cite{Holmegaard:PRL102:023001, Filsinger:JCP131:064309,
   Trippel:PRA89:051401R, Trippel:PRL114:103003}.

\begin{figure}[t]%
   \centering
   \includegraphics[width=\linewidth]{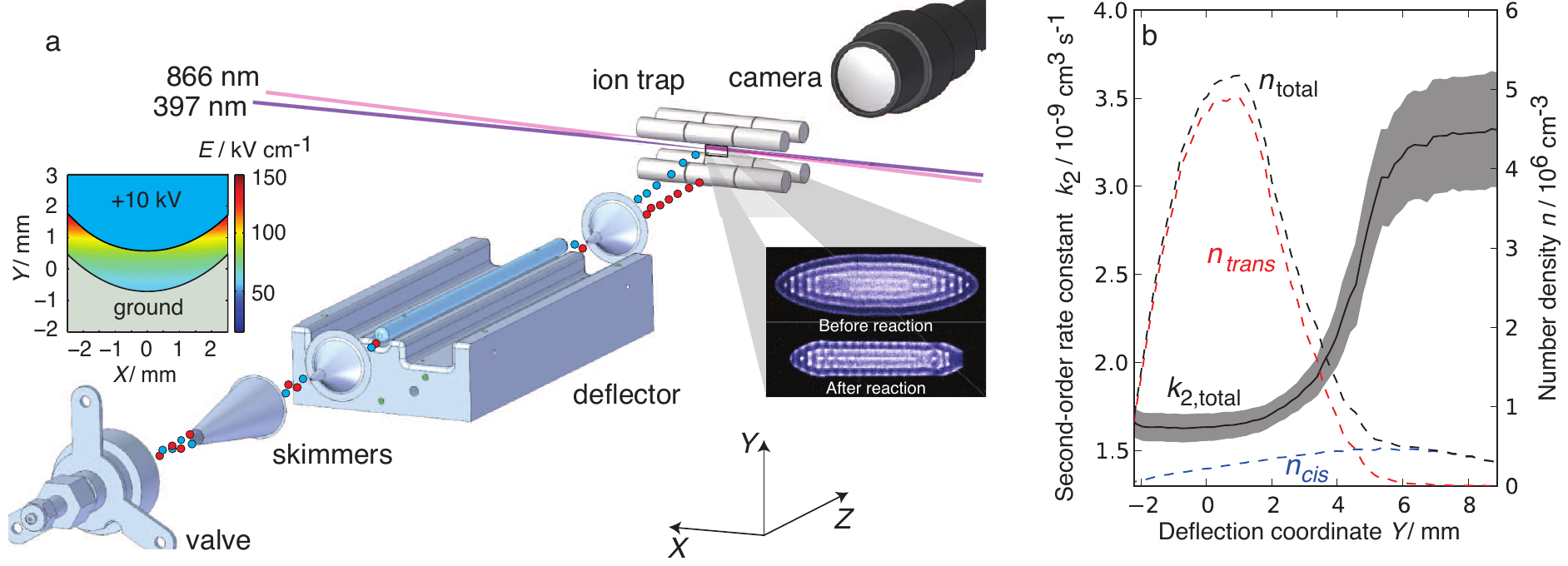}
   \caption{Conformer-selected chemistry investigated through reactive collisions between \trans and
      \cis 3-aminophenol with laser-cooled Ca$^+$ ions in a Coulomb crystal. (a) Experimental setup
      combining the electrostatic deflector with the quadrupole ion trap containing the Coulomb
      crystal. Tilting of the molecular beam relative to the trap allows to select the conformer
      incident on the crystal. (b) Measured second-order rate constant as a function of deflection
      coordinate. Shown as dashed lines are the relative intensities for each conformer at that
      position. Reproduced from reference~\citealp{Chang:Science342:98}. Reprinted with permission from AAAS.}
   \label{fig:applications:reactions}
\end{figure}

Recently, the conformer separation with static electric fields has been used to study conformer
resolved reactions in scattering experiments of conformationally selected neutral molecules with
trapped cold atomic ions~\cite{Chang:Science342:98, Roesch:JCP140:124202}. Specifically,
conformer-selected molecular beams of \textit{cis}- and \textit{trans}-3-aminophenol were collided
with a stationary target of laser-cooled Ca$^+$ ions and the reaction rate measured for both
conformers, this experimental setup is shown in~\autoref{fig:applications:reactions}\,a.
By measuring the rate constants at different points in the deflected molecular beam, shown
in~\autoref{fig:applications:reactions}\,b, the conformer-specific rate constants were extracted and
the \textit{cis} conformer found to have a rate constant approximately twice that of the
\textit{trans} species. Measurement of the number densities in the molecular beam furthermore
allowed
the extraction of absolute values for the second order rate constant for each conformer. These
findings were explained with conformer-specific differences in the long range interaction potential
of the molecule and ion, stemming from the conformation-dependent electrostatic properties of the
molecules. This experiment nicely demonstrates the fact that these conformers exhibit distinct
chemical properties, \ie, are separate molecular species as outlined in \autoref{sec:introduction}.

This proof-of-principle experiment shows the detailed level of quantitative information to be gained
from conformer-resolved gas-phase experiments, and several extensions to this setup are currently
being planned or implemented. The production of molecular Coulomb
crystals~\cite{Willitsch:IRPC31:175}, sympathetically-cooled from collisions with laser-cooled
atoms, opens up possibilities for the study of ion-molecule reactions, with the possibility to
extract absolute densities and rate constants for individual conformer reactions. These experiments
can add further selectivity by employing narrowband resonant ionisation schemes for the production
of ions, yielding rotational and vibrational state specificity for the ionic reaction
partner~\cite{Tong:PRL105:143001}.

\subsection{Cluster selection: From gas-phase to bulk chemistry}
There is substantial interest in the study of molecular cluster systems, which could bridge the gap
between the gas-phase, where highly detailed investigations of the intrinsic molecular properties
are feasible, and the condensed phase, where the majority of chemistry occurs. Furthermore, clusters
allow to sequentially and systematically add solvent molecules to the molecular species of interest.
This aids approaches to unravel the effects of solvation, for instance, on electronic structure.
Significant literature is available on the study of ionic clusters, where size-selection is trivial
using mass-spectroscopic techniques~\cite{Verlet:CSR37:505, Fujii:IRPC32:266}. This is not the case
for neutral cluster experiments, where experimenters have so far primarily relied on the
spectroscopic identification of formed clusters, and only few experimental techniques can physically
separate clusters, such as scattering from a secondary Helium beam~\cite{Buck:PRL52:109,
   Goerke:ZPD19:137, Buck:CR100:3863}. The application of the alternating-gradient focuser to the
separation of cluster systems has recently been demonstrated in a proof-of-principle experiment.
Here, benzonitrile-argon clusters could be separated partially from benzonitrile due to the
different masses, but nearly identical dipole moments, of the complexes~\cite{Putzke:JCP137:104310}.
Use of the simple electrostatic deflector offers a widely applicable method of producing pure
clusters beams. It relies on different cluster stoichiometries possessing differing dipole moments
and/or different masses, too, allowing their separation according to the $m/$\mueff ratio. In
molecular beam experiments this enables the separation of small clusters from remaining monomers or
other cluster stoichiometries in the beam. Furthermore, this method is in principle also sensitive
to the geometric structure of the cluster through the dipole moment dependence and should enable the
separation of different structural cluster isomers, \eg, the different geometries of the water
hexamer~\cite{Perez:Science336:897}.
\begin{figure}[t]%
   \centering%
   \includegraphics[width=0.6\linewidth]{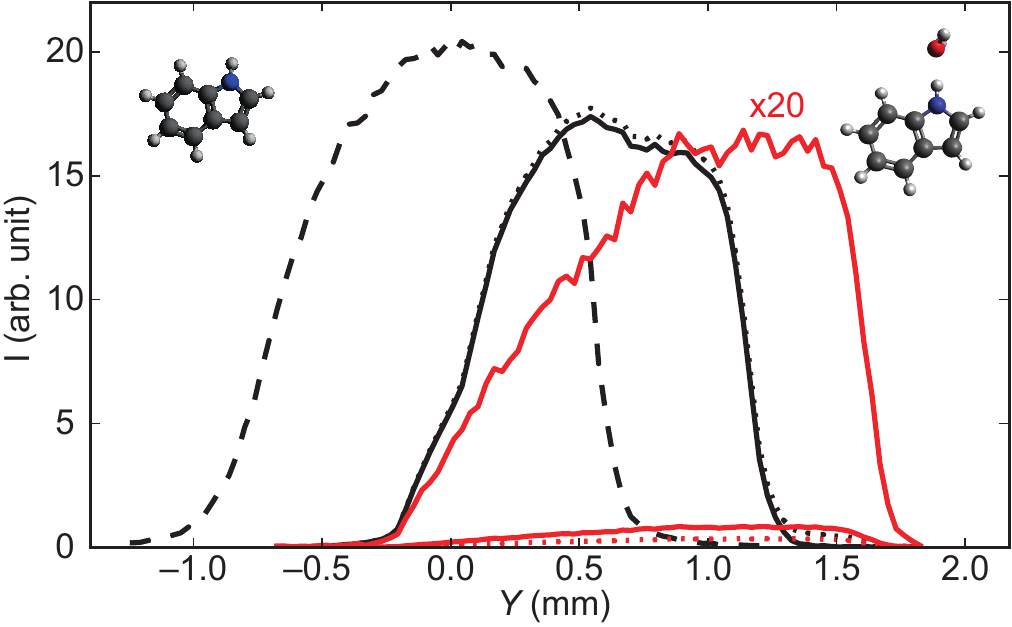}%
   \caption{Deflection profiles from the co-expansion of indole and water, leading to the formation
      of \indww{n} clusters. The dashed line shows the field-free molecular beam profile. Dotted
      lines indicate measured signal at the mass channels corresponding to singly charged bare
      indole (black) and \indw (red). Solid lines show the molecular beam density, corrected for
      fragmentation of the \indw cluster. At vertical positions in the range $\sim$1.4--1.8~mm a
      pure \indw cluster beam is obtained.}
   \label{fig:applications:clusters}
\end{figure}

While it is not \textit{a priori} clear if the deflection technique is applicable to larger, floppy
molecular systems such as clusters, containing such high densities of states and correspondingly
complex Stark energy curves with multiple avoided and real crossings, its applicability to small
cluster systems has recently been demonstrated in the study of \indww{n}
clusters~\cite{Trippel:PRA86:033202}. Here, indole and water were co-expanded in a helium-seeded
molecular beam, leading to the formation of a \textit{cluster soup}, containing the respective
monomers and different clusters of the type \indww{n}. The produced beam was directed into a 15~cm
long deflector with field strengths on the order 120~kV/cm. Spatial profiles were recorded
species-selectively for indole and water monomers as well as the \indw clusters using
resonance-enhanced multi-photon ionization (REMPI)~\cite{Trippel:PRA86:033202}. While in the absence
of a deflection field all species are spatially overlapped, application of strong fields broadens
the observed distributions, due to the different \mueff of occupied rotational states. The \indw
cluster possess the largest dipole moment ($\mu=4.4$~D)~\cite{Kang:JCP122:174301} and is deflected
significantly more than indole ($\mu=1.96$~D). Similarly, water molecules ($\mu=1.86$~D) and higher
order clusters such as \indww{2} ($\mu=2.63$~D) are deflected significantly
less~\cite{Trippel:PRA86:033202}. \autoref{fig:applications:clusters} shows an updated experiment of
the cluster separation with a colder beam than in the original experiment. Here, strong-field
ionization mass-spectrometry was used in a high-repetition-rate
experiment~\cite{Trippel:MP111:1738}. This leads to the production of a pure \indw sample, at
spatial positions $1.4<Y<1.8$~mm. The spatial dispersion of rotational quantum states furthermore
means that at the edge of the deflected profile only the lowest $\sim\!290$ rotational quantum
states of \indw are present, compared to $\sim$4600 states in the original
beam~\cite{Trippel:PRA86:033202}. An electrostatic deflector, therefore, not only produces a pure
cluster beam, but also yields lower effective rotational temperatures than can be obtained from
supersonic expansion alone.

\subsection{Separation of individual quantum states}
\label{sec:quantumstates}
The production of single quantum states samples is a routine technique for low-field-seeking (lfs)
states in small molecules, first developed in the 1950s~\cite{Bennewitz:ZP139:489,
   Bennewitz:ZP141:6} and relying on multipole focusers and static fields to guide and select lfs
quantum states~\cite{Reuss:StateSelection, Bethlem:PRL83:1558, Meerakker:CR112:2012,
   Brouard:CSR:2014}. These were crucial for the development of the first
MASER~\cite{Gordon:PR95:282,Gordon:PR99:1264}, and later on in the first inelastic and reactive
state-specific scattering experiments~\cite{Bennewitz:ZP177:84, Brooks:JCP45:3449,
   Beuhler:JACS88:5331}. The inability to create a field maximum in free space, however, restricts
these methodologies to lfs states, and, therefore, cannot produce isolated ground state samples of
larger molecules, for which all low-lying quantum states are hfs. The Stark deflection technique
presented here does allow the separation of hfs quantum-states through the quantum-state dependence
of the effective dipole moment \mueff. It does, however, not provide guiding or focusing of the
selected molecules, meaning that devices are typically restricted to shorter lengths to ensure
sufficient densities in the interaction region. Dynamic field techniques, such as
alternating-gradient focusers~\cite{Putzke:PCCP13:18962}, have demonstrated the ability to provide
guiding for state-selected hfs states, albeit at the expense of experimental simplicity, making it
extremely challenging to incorporate these devices into existing experimental setups. To achieve a
full separation of quantum-states with the Stark deflector, the difference in \mueff for
neighbouring states must be sufficiently large to separate them. This is the case for small
molecules with correspondingly large rotational constants and rotational energy level separation.
Specifically, single-quantum states of OCS~\cite{Nielsen:PCCP13:18971, Trippel:PRA89:051401R} and
water~\cite{Horke:ACIE53:11965} have been produced in the gas-phase. As electrostatic deflection
does not change the population of quantum-states in the molecular beam, but rather disperses it, a
large initial population of the lowest quantum states, \ie, a low rotational temperature, is crucial
to these experiments.

The first demonstration of single-quantum states produced with the electrostatic deflector was for
OCS molecules~\cite{Nielsen:PCCP13:18971}. OCS was produced \textit{via} supersonic expansion in a
pulsed valve, producing a molecular beam in the electronic and vibrational ground state
($X^1\Sigma^+ \ket{00^00}$) and with a rotational temperature $<1$~K. The beam is then dispersed by
a 15~cm long electrostatic deflector, leading to a spatial separation of rotational quantum states
due to their different effective dipole moments.
\begin{figure}[t]%
   \centering
   \includegraphics[width=0.8\linewidth]{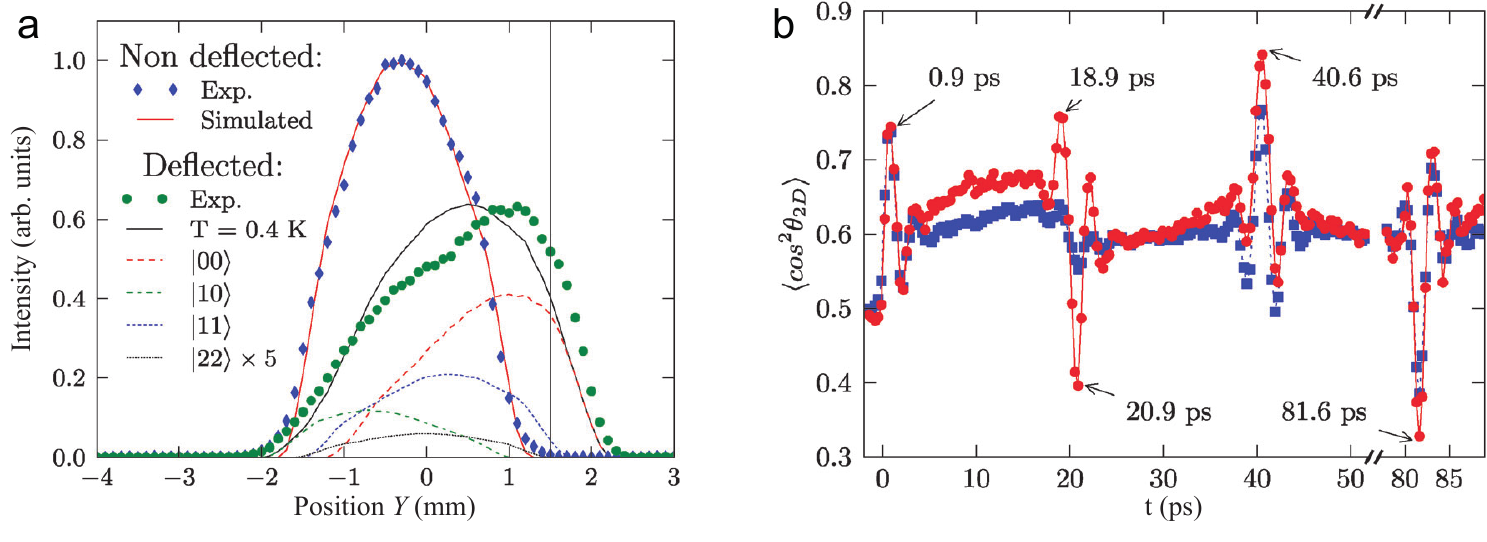}
   \caption{Quantum-state separation of OCS. (a) Molecular beam intensity profiles (data points) and
      simulations for individual quantum states (dashed lines) and for a convolution at 0.4~K
      rotational temperature (solid line). (b) Non-adiabatic alignment dynamics for OCS, as
      characterised by the time-dependent degree of alignment
      $\langle\cos^2\theta_{\text{2D}}\rangle$, for the direct beam (blue) or the deflected beam at
      1.5~mm (red). Reproduced from reference~\citealp{Nielsen:PCCP13:18971} with permission of The Royal
       Society of Chemistry.}
   \label{fig:applications:OCS}
\end{figure}
This is shown in \autoref{fig:applications:OCS}\,a, along with corresponding trajectory simulations
indicating the contributions of different quantum states and the overall ensemble temperature,
introduced in \autoref{sec:theory-state-selection}~\cite{Chang:CPC185:339}. This experiment did not
employ quantum-state resolved detection, but the separation of quantum states could directly be
inferred from comparison to theory as well as a variant of rotational-coherence spectroscopy. The
comparison with the trajectory simulations indicate that at the position $Y=1.5$~mm at least $85\%$
of molecules are in the absolute ground state $\ket{00}$. An impulsive-alignment experiment shows
clear quarter-revival features for the deflected beam, \autoref{fig:applications:OCS}\,b, which are
the direct result of an ensemble of defined parity. Comparison of the phase and amplitude of the
revivals with quantum-dynamics calculations indicate a purity of 92\% of the $\ket{00}$ state in the
deflected beam~\cite{Nielsen:PCCP13:18971}.

\begin{figure}[t]
   \centering
   \includegraphics[width=0.8\linewidth]{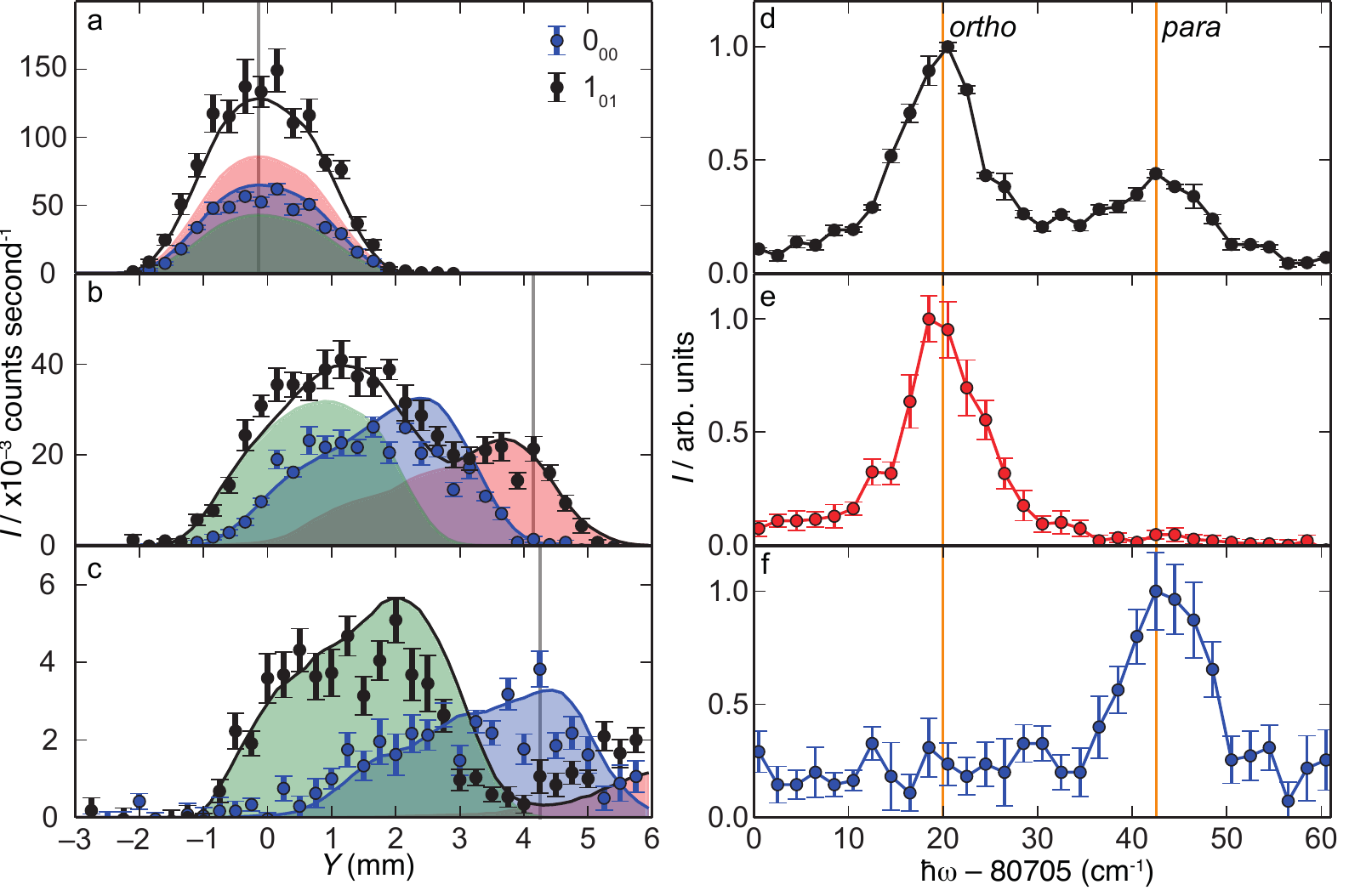}
   \caption{Separation of the nuclear-spin isomers of water with the electrostatic deflector. On the
      left are spatial profiles measured using REMPI (data points) and the corresponding simulations
      (solid lines), the shading indicates the $\ket{0_{00}}$ (green), $\ket{1_{01}0}$ (blue) and
      $\ket{1_{01}1}$ (red) states. Profiles are shown for a Neon expansion at (a) 0~kV and (b)
      15~kV, and for Argon at 15~kV (c). Corresponding REPMI spectra are shown in d-f, recorded at
      the position marked by the grey lines in the spatial profile.}
   \label{fig:applications:water}
\end{figure}

The production of single-quantum-state samples of large molecules in hfs states has also been
demonstrated using alternating-gradient focusing~\cite{Putzke:PCCP13:18962}. These setups allow one
to tune the transmitted \mueff$/m$ range by changing the electrode switching frequency and voltages.
This allowed the production of pure absolute ground state samples of benzonitrile and a
$\mu/\Delta\mu$ resolution $\sim20$ is achieved~\cite{Putzke:PCCP13:18962}. The application of the
deflector for the production of single-quantum-state samples of small asymmetric top molecules was
recently demonstrated with the first full separation of the nuclear spin states, \para and \ortho,
of water~\cite{Horke:ACIE53:11965}. Here, the two absolute ground states of the nuclear-spin isomers
could be spatially separated and pure samples of either obtained. These isomers differ only in the
relative orientation of the hydrogen nuclear spins, leading to a symmetric (\textit{para}) and an
antisymmetric (\textit{ortho}) spin wavefunction. The symmetrization postulate of quantum mechanics
requires the overall molecular wavefunction to be antisymmetric, and, therefore, the nuclear spin
states reside in different rotational states. A cold water beam was produced in a supersonic
expansion, leading to rotational temperatures $<7$~K, but nuclear-spin populations that are frozen
at the room-temperature limit of 1:3 \textit{para}- to \textit{ortho}-water. The absolute ground
states $\ket{J_{K_aK_c}}$ of these two species are $\ket{0_{00}}$ for \textit{para}- and
$\ket{1_{01}}$ for \textit{ortho}-water, respectively. In the presence of an electric field the $M$
degeneracy will be lifted, splitting the \textit{ortho}-water state into $M=0$ and $M=\pm1$
components, with distinctive \mueff. These three populated states can all be separated with the
electrostatic deflector, as shown in \autoref{fig:applications:water}~\cite{Horke:ACIE53:11965}.
The $\ket{1_{01}1}$ has the largest \mueff and, therefore, undergoes the largest deflection; this
can be separated from all other quantum-states, \autoref{fig:applications:water}\,b and e. To
produce pure \textit{para}-water the deflection is increased through the use of a heavier backing
gas and, therefore, slower molecular beam, which increases the interaction time with the electric
field. This deflects all $\ket{1_{01}1}$ population out of the interaction region and produces a
pure $\ket{0_{00}}$ sample, \autoref{fig:applications:water}\,c and f. A sample of $\ket{1_{01}0}$
can be accessed with this configuration at the undeflected beam position, as all other quantum
states have been deflected away. These separated nuclear-spin isomers of water could allow
significant applications in astrophysics and -chemistry as well as novel NMR imaging techniques.
Furthermore, it highlights the general applicability of the deflector to access the absolute ground
state of neutral molecules, due to its sensitivity to hfs species. Moreover, it adds further support
for the concept of molecular structure and species brought forward in \autoref{sec:introduction}.

\subsection{Cold molecular samples through rotational-state dispersion}
\label{sec:alignment}
The selection of a single rotational quantum state of a molecule gets increasingly harder as the
molecular structure gets larger and, therefore, the density of states increases. However, the
electrostatic deflection technique, nonetheless, leads to a dispersion of the rotational states by
their effective dipole moment, equivalent to the dispersion of white light by a prism. As the lower
rotational states have larger \mueff, \autoref{fig:sel:Stark_curves}, these are deflected the most,
leading to a spatially dispersed rotational state distribution. This is not an active cooling
process as the population distribution among the rotational states remains unchanged, but rather
allows the probing of a lower effective rotational temperature by excluding high $J$ states from the
probe volume. Therefore, this technique will reduce the number density available and requires a high
initial population of the lowest rotational states, necessitating a cold initial molecular beam.

\begin{figure}[t]
   \centering
   \includegraphics[width=0.5\linewidth]{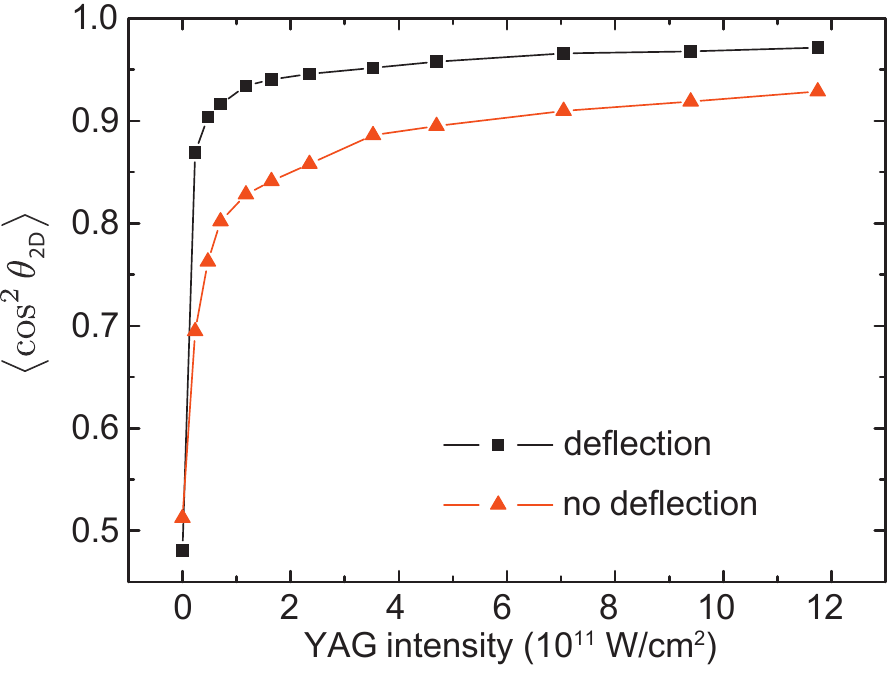}
   \caption{Degree of alignment observed for iodobenzene molecules for different alignment laser
      intensities, recorded with a direct molecular beam (red) and with the coldest part from an
      electrostatically dispersed beam (black). Reprinted with permission from
      reference~\citealp{Filsinger:JCP131:064309}. Copyright 2009, AIP Publishing LLC.}
   \label{fig:applications:alignment}
\end{figure}

One important application for these internally-ultra-cold samples are molecular alignment and
orientation experiments~\cite{Kumarappan:JCP125:194309}. Especially in the regime of long-pulse
``adiabatic'' alignment~\cite{Stapelfeldt:RMP75:543, Trippel:MP111:1738, Trippel:PRA89:051401R},
these experiments benefit from the lower-$J$ state distribution in the deflected part of the
molecular beam and a higher degree of alignment can be reached, as shown in
\autoref{fig:applications:alignment}~\cite{Holmegaard:PRL102:023001, Filsinger:JCP131:064309}.
This has been demonstrated for several small aromatic molecules~\cite{Holmegaard:PRL102:023001,
   Nevo:PCCP11:9912, Holmegaard:NatPhys6:428} and, furthermore, has been extended to kilohertz
repetition rates~\cite{Trippel:MP111:1738}. When combined with the single-quantum-state selection
achieved for OCS, this allows for the observation and exploitation of specific quantum-mechanical
effects. For example, Raman-type couplings during the switch-on of a strong laser field lead to the
creation of a pendular-state wavepacket, forming a strongly driven quantum-pendulum, observable
through the degree of molecular alignment~\cite{Trippel:PRA89:051401R}. A similar effect of coherent
superposition of states within strong fields can be utilized to achieve field-free orientation of
molecules, as the coherent superposition persists, and, therefore, revives, even after the laser
pulse has been turned off~\cite{Trippel:PRL114:103003}.

\subsection{Imaging experiments using novel light sources}
The advance of new coherent x-ray light sources providing ultrashort pulses, such as free-electron
lasers and high-harmonic generation techniques, offer the opportunity to record nuclear dynamics at
the ultrashort timescales at which atomic motion occurs, allowing the recording of \emph{molecular
   movies}. Such sources are, however, highly reliant on the controlled and reproducible delivery of
samples into the interaction region. For x-ray diffraction techniques on isolated molecules, it is
furthermore highly desirable to control the spatial alignment and orientation of molecules. A
further complication when studying gas-phase species with ionizing radiation is the typically low
number density of molecules, especially when compared to the carrier gas of the molecular expansion,
which is on the order of $10^4$ as dense.

Several of these experimental difficulties can be addressed using the electrostatic deflection
technique. As outlined above, the dispersion of rotational quantum-states allows the production of
colder ensembles of molecules, which can be used to achieve unprecedented degrees of alignment.
Furthermore, as the monoatomic carrier gas passes through the deflector unaffected, the molecules of
interest are actually separated from the stream of backing gas, potentially significantly reducing the
background signal observed in X-ray experiments, such as single molecule diffractive
imaging~\cite{Kuepper:PRL112:083002} or photoelectron diffraction in aligned molecules following
core-shell excitation with x-ray pulses~\cite{Boll:PRA88:061402,Rolles:JPB47:124035}. Finally, as
these x-ray experiments are inherently non species-selective, a pure sample in the interaction
region is required. Using the electrostatic species separation, this is now feasible for large
complex molecules containing several conformers, and even weakly-bound cluster systems.

\begin{figure}[t]
   \centering
   \includegraphics[width=1\linewidth]{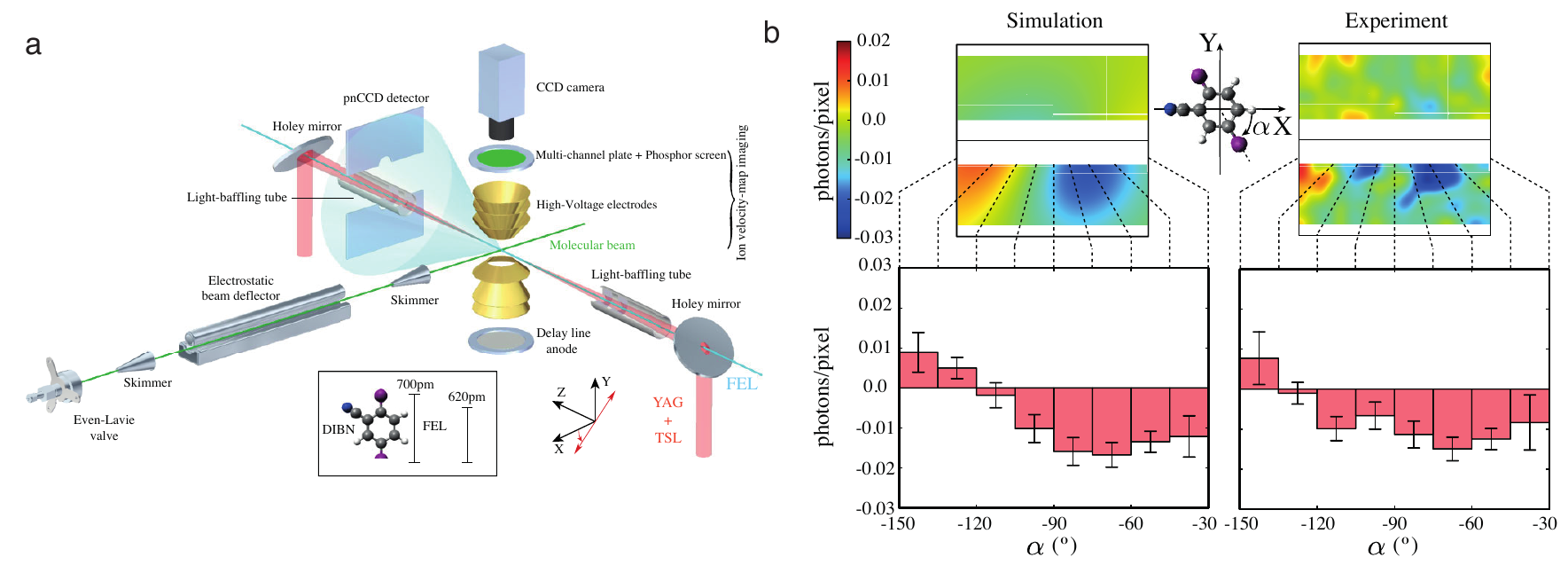}
   \caption{X-ray diffraction from aligned molecules. (a) Experimental setup with high-pressure
      valve and electrostatic deflector for the production of ultracold molecular samples. The
      interaction takes place inside a VMI spectrometer to monitor the degree of alignment using ion
      imaging. X-ray diffraction patters are recorded on a pnCCD detector. (b) Simulated and
      experimental difference diffraction patterns and angular histograms for diffraction of 2~keV
      photons from 2,5-diiodobenzonitrile. Reprinted with permission from reference~\citealp{Kuepper:PRL112:083002}.
      Copyright 2014 by the American Physical Society.}
   \label{fig:applications:diffraction}
\end{figure}

A recent proof-of-principle experiment has demonstrated this approach with the first
x-ray-diffraction measurement from aligned gas-phase molecules using a free-electron
laser~\cite{Kuepper:PRL112:083002, Stern:FD171:393}. These experiments can provide direct access to
structural information. 2,5-diiodobenzonitrile molecules were supersonically expanded into vacuum
and subsequently dispersed by the electrostatic deflector to select a colder sample. These molecules
were adiabatically aligned with a nanosecond laser pulse and irradiated with bright x-ray pulses
from the Linac Coherent Light Source, whose direct scattering was recorded. A typical experimental
setup is shown in \autoref{fig:applications:diffraction}\,a. The alignment laser and x-ray pulses
are combined using a holey mirror and propagate collinearly, crossing the state-selected molecular
beam at right angles. The interaction takes place within a velocity-map imaging spectrometer, where ion
imaging is utilized to quantify the degree of alignment of the target molecules. Scattered x-rays
are recorded on a pnCCD detector, with a central gap to let the main
beams pass through.
Recorded and simulated difference diffraction patterns of aligned and randomly oriented molecules
are shown in \autoref{fig:applications:diffraction}\,b, along with the corresponding angular
distribution histograms. In this proof-of-principle experiment the dominant scatterer are the heavy
iodine atoms, and the recorded diffraction images allowed the authors to infer the
iodine-iodine distance in 2,5-diiodobenzonitrile, in good agreement with literature and simulated values.
These results demonstrate the feasibility of measuring
 structural information from diffraction images of single molecules, aligned in
the gas-phase. The very strong degree of alignment is essential in experiments aimed at extracting
molecular frame 3D structural data, and it is the electrostatic deflection technique that provides the
necessary ultracold molecular samples for this.

\section{Outlook}
\label{sec:outlook}
\subsection{Separation of very large molecules}
\label{sec:outlook:very-large}
A variety of new and exciting applications can be envisioned for the electrostatic deflector. The
separation of conformers and quantum states holds in general. In practice, it allows the separation
of molecular species that differ sufficiently with respect to their effective dipole moments in
strong electric fields. The influence of very strong electric fields on both the orientation of the
dipole as well as structural changes have been investigated theoretically for large molecules such
as polyalanines~\cite{Calvo:BiophysJ95:18} or the enzyme lysozyme~\cite{Abrikossov:thesis:2011}. Our
simulations show that the molecular dipole moment becomes oriented in static electric fields for
field strengths on the order of 100~kV/cm. Structural transformations that distort the molecular
structure only appear at significantly higher field strengths~\cite{Calvo:BiophysJ95:18,
   Abrikossov:thesis:2011} and these effects do not play a significant role for the experimental
conditions in the deflector. Thus, the dispersion of beams of such molecules is feasible and will
allow for a, at least partial, separation of different species. Especially for the helical
structures of peptides, in which all amino-acid dipoles are parallel, a very strong deflection and
the separation from other folding motifs can be expected.

\subsection{Separating the enantiomers of chiral molecules}
\label{sec:outlook:chiral}
Preparing molecules in a certain quantum state could be beneficial for precision experiments to test
fundamental physics. The CPT theorem states that the laws of physics are invariant under the
simultaneous transformations of charge conjugation (C), parity transformation (P), and time reversal
(T). Although CPT is conserved, parity (P) might not be conserved in processes involving the weak
force~\cite{Lee:PR104:254, Wu:PR105:1413}. This leads to parity-violation effects, \eg, between the
two enantiomers of chiral molecules~\cite{Darquie:Chirality22:870, Quack:ProgressTheoChemPhys26:47}.

The potential energy surface of a chiral molecule shows two minima separated by a barrier. These
minima may be associated with the left ‘L’ and right ‘R’ enantiomers. When the interconversion
barrier is high, the right and left state can be, to a good approximation, considered as energy
eigenstates. These eigenstates are degenerate within quantum mechanics in the absence of the weak
force. In the presence of the weak force however, a small parity violation energy difference is
expected between the ground states and excited states of the enantiomers. Therefore, right- and
left-handed molecules are not exact mirror images of each other. The energy differences of
enantiomers is predicted in the femtojoule to picojoule per mole range~\cite{Quack:ACIE41:4618}.
This small difference is the reason why parity violation effects in chiral molecules have so far
never been observed experimentally. The energy differences between two enantiomers might be
accessible in high-resolution laser- or microwave-spectroscopy~\cite{Daussy:PRL83:1554,
   Darquie:Chirality22:870, Medcraft:ACIE53:11656}. In order to study parity violation effects in
molecules it would be highly beneficial to study both enantiomers individually in order to determine
the differences in the energy eigenstates. The electrostatic deflector might provide a useful tool
in order to separate both enantiomers, online, for future high precision experiments as well as for
further investigations of their individual properties, including chemical
reactions~\cite{Chang:Science342:98}.

For such an experiment, one could, for instance, start with two enantiomers of a molecule in the
absolute ground state. This would be achieved, for a molecule such as CHBrClF, with the smallest
rotational constant on the order of 1~GHz, at a rotational temperature of about 10~mK or using an
appropriate state-selection experiment (see, \eg, \autoref{sec:quantumstates}). Starting from both
enantiomers in their ground state, one could, using resonant microwaves, prepare one of the
enantiomers in a rotationally excited state before it enters the deflector, while the second
enantiomer stays in its ground state. The two enantiomers will then be deflected differently due to
the different space-fixed dipole moments \mueff of the two states. The method to excite only one
enantiomer is conceptionally similar to the enantiomer-specific detection of chiral molecules by
microwave spectroscopy~\cite{Patterson:Nature497:475}.

\begin{figure}
   \centering
   \includegraphics[width=1.0\linewidth]{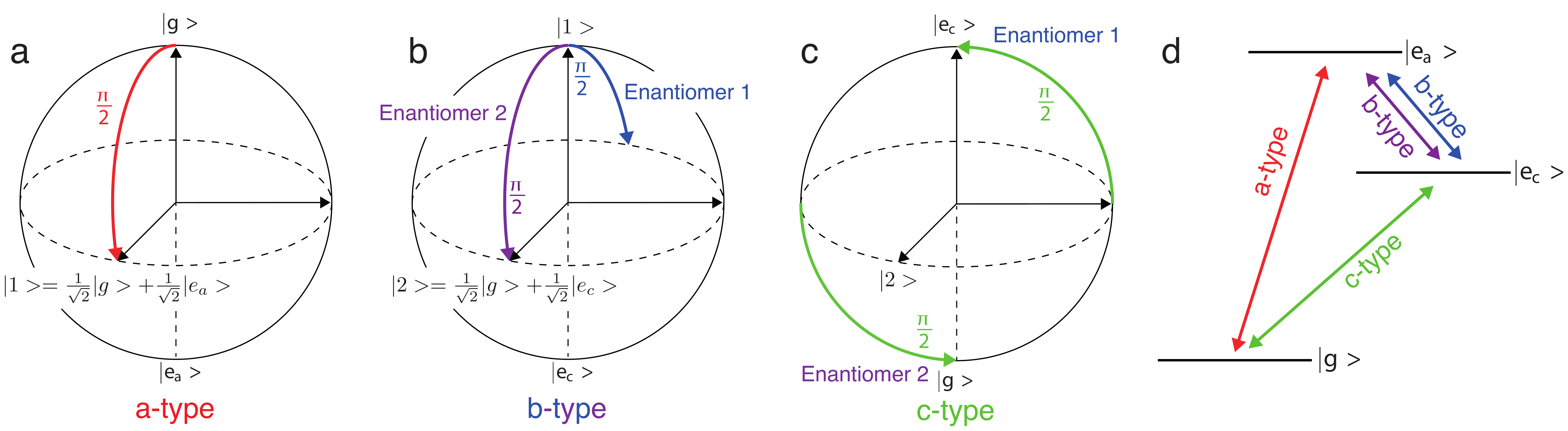}
   \caption{Illustration of the excitation scheme to transfer one enantiomer to the excited
     state. a) Bloch-sphere of the two-level system consisting of the states \ket{g} and
     \ket{e_a}. The $\pi/2$ pulse of the corresponding $a$-type transition, that transforms both
     enantiomers to the pure state $\ket{1}=1/\sqrt{2}\ket{g}+1/\sqrt{2}\ket{e_a}$, is indicated by
     the red quarter circle. b) Bloch-sphere of the two-level system consisting of the states \ket{1}
     and \ket{e_c}. The $\pi/2$ pulse of the corresponding $b$-type transitions is indicated by the blue
     quarter circle for enantiomer 1 and the purple quarter circle for enantiomer 2. The phase
     between the applied a- and b-type transitions has to be such, that the resulting state
     $\ket{2}=1/\sqrt{2}\ket{g}+1/\sqrt{2}\ket{e_c}$ has no contribution from \ket{e_a}. c)
     Bloch-sphere of the two-level system consisting of the states \ket{g} and \ket{e_c}. The $\pi/2$
     pulse of the corresponding $c$-type transition, that transfers the system from the state \ket{2} to
     the excited state \ket{e_c} for enantiomer 1 and to the ground state \ket{g} for enantiomer 2,
     is indicated by the green quarter circles. d) Level diagram of the involved states and
     transitions.}
   \label{fig:pi-pulses}
\end{figure}

Asymmetric top molecules have three inequivalent principal axes of inertia, $a,b,c$, with rotational
constants $A$, $B$, and $C$, which describe the rotational energy levels. The corresponding
magnitudes of the dipole moment components, $|\mu_a|,|\mu_b|,|\mu_c|$, determine the strengths of
transition between the rotational energy levels of such a molecule, while the sign of
$\mu_a\cdot\mu_b\cdot\mu_c$ is directly connected to its chirality. For simplicity, both enantiomers
are assumed to be in their absolute ground state \ket{g} initially. The two enantiomers are exposed
to a resonant $\pi/2$ pulse with polarization, for instance, along the $a$ axis, \ie, corresponding
to an $a$-type transition. This creates a superposition state,
$\ket{1}=1/\sqrt{2}\ket{g}+1/\sqrt{2}\ket{e_a}$, between the ground \ket{g} and some excited state
\ket{e_a} as indicated in~\autoref{fig:pi-pulses}~a by the red quarter circle and arrow.

A second $\pi/2$ laser pulse with a polarization perpendicular to $a$, \eg, $b$, which is resonant
to a $b$-type transition, mixes in components of a second excited state \ket{e_c} with complex
phases proportional to $\mu_b$ (purple and blue lines). The phase between the applied $a$-type
transition that mixes \ket{g} and \ket{e_a} and the $b$-type transition that mixes \ket{e_a} and
\ket{e_c}, has to be ideally chosen such that the new state \ket{2} is given by
$\ket{2}=(1/\sqrt{2}\ket{g}+1/\sqrt{2}\ket{e_a})-(1/\sqrt{2}\ket{e_a}-1/\sqrt{2}\ket{e_c})=1/\sqrt{2}\ket{g}+1/\sqrt{2}\ket{e_c}$.
The resulting state \ket{2} has no contribution from \ket{e_a} and the two pulses have now created a
superposition of the two states \ket{g} and \ket{e_c} oscillating along the $c$ axis, which is
perpendicular to $a$ and $b$. The complex Rabi frequency, describing an electric dipole transition
between rotational states of a molecule, has opposite sign for the two
enantiomers~\cite{Jacob:JCP137:044313}. Therefore, the oscillation along the $c$ axis has a phase
shift of $\pi$ for the two enantiomers (blue quarter circle for the first and purple quarter circles
for the second enantiomer in~\autoref{fig:pi-pulses}~b). Adding a third $c$-polarized $\pi/2$ pulse
connecting \ket{g} and \ket{e_c}, with a controlled phase relative to the first two pulses, it is
possible to transfer the population of one enantiomer to the excited state and the population of
the second enantiomer back to the ground state. This is indicated in \autoref{fig:pi-pulses}~c by
green quarter circles and arrows. Finally, when traversing an electric deflector, the two
enantiomers in the molecular beam will be spatially dispersed by the electrostatic deflector due to
the different effective dipole moments of the two states they reside in. Enantiomeric pure samples
are obtained by making use of skimmers properly positioned in the molecular beam and are thus
available for applications, \eg, precision-spectroscopy experiments.

\subsection{Nonadiabatic effects}
\label{sec:outlook:nonadiabatic-effects}
The complex energy level structure of molecules in external fields provides a serious challenge for
their theoretical description~\cite{AbdElRahim:JPCA109:8507, Haertelt:JCP128:224313,
   Filsinger:JCP131:064309, Omiste:PCCP13:18815, Nielsen:PRL108:193001, Hansen:JCP139:234313}.
Especially for the general case of an asymmetric top a high density of states and low symmetry of
the system~\cite{Chang:CPC185:339} leads to many adiabatically avoided crossings. Studies of the
nonadiabaticity in the network of dense crossings is a challenging but intriguing problem. The
investigation of the resulting nonadiabatic dynamics in the interaction of a molecule with a strong
electric field might be accessible through the electric deflector. In its inhomogeneous strong
electric field the molecules are exploring the energy landscape of their pendular states. The
passing speed of the system at an avoided crossing, and, therefore, its adiabaticity is dependent on
the rate of change of the electric field strength. By controlling the rate of change of the applied
voltages for a defined electrode geometry, it is possible to control the speed of passing through
the avoided crossings. This can be achieved through high-voltage switching using experimentally
well-defined rise and fall times. Moreover, the deflector can be used as a quantum state detector.
In a first experiment the quantum state properties and populations in the molecular beam are
characterized with the electric deflector. In a second experiment the molecules are exposed to a
time-dependent electric field before they enter the deflector. The rise time of the electric field
is chosen such that the molecules pass an avoided crossing non-adiabatically. The fall time of the
electric field should ideally be slow enough to recross it adiabatically. Such a switching procedure
ensures that the molecules only traverse the avoided crossing non-adiabatically once. The modified
quantum-state distribution is again characterized by the electric deflector. Multiple field
gradients can be explored to increase the obtained amount of information on the avoided crossing.
Comparison with the quantum-state distribution without switching the electric field yields the
population transfer distribution through the avoided crossing and allows to unravel the physics of
non adiabatic behaviour at individual and connected avoided crossings. For typical avoided crossings
a rise time of the electric field on the order of $10^{18}~\text{V}/(\text{m}\,\text{s})$ (\eg,
$1000~\text{kV}/\text{cm}$ in 100~ps) is required in order to become impulsive. This is about three
orders of magnitude faster than possible with current technology, but the necessary technologies
might become available in the future.

\subsection{Cyclic deflectors}
\label{sec:outlook:cyclic}
The quantum state resolution behind a deflecting device can be increased by changing the shape of
the electrodes \cite{DeNijs:PCCP13:19052}, extending it or by using several deflectors behind each
other. Increasing the length of a single straight deflector will eventually result in the molecules
crashing into its electrodes~\cite{Filsinger:ACIE48:6900}. Alternatively a curved deflector
electrode could be used. Assuming a favorable deflection of 3~mm after a 15~cm long deflector this
would result in a radius of approximately 4~m for such a device, allowing for very strong
quantum-state discrimination. We point out that the molecules are not actually trapped, since they
are in hfs states and, therefore, are always attracted by the electrodes of the deflector. This has
the consequence that such a setup is not a storage ring. Nevertheless, the molecules will traverse
the device for a significant amount of time on metastable trajectories. For typical beam conditions
as described in \autoref{sec:applications}, we estimate a number density of $10^{4}/\text{cm}^3$ in
a very pure single-quantum-state sample after one round trip in such a device.

\subsection{Merged-beam reactive scattering}
\label{sec:outlook:merged-beam-reactions}
Extending the ion-molecule reactions described in \autoref{sec:application:conformer-selection},
applications for crossed- and merged-beam reactive-scattering experiments with large molecules are
envisioned, similar to current experiments with atoms and small
molecules~\cite{Kirste:Science338:1060, Henson:Science338:234, Jankunas:JPCA118:3875}. Using curved
deflector electrodes (see \autoref{sec:outlook:cyclic}), two velocity-matched beams of polar
molecules can be merged in virtually any, lfs or hfs, single or small set of quantum states. Due to
the low relative velocities, and the correspondingly small number of partial waves contributing to
the reaction, possible scattering resonances~\cite{Yang:ARPC58:433, Henson:Science338:234,
   Chefdeville:Science341:1094} may become visible even in the reactions of complex molecules.

In addition a movable skimmer behind the deflector could serve as a selector for molecules in
specific quantum states. This results in a molecular beam of selected molecular species in selected
quantum-states, which can be exploited for reactive scattering experiments. Furthermore, a deflector
in the outgoing beam could even disperse and analyse the product states and species distributions.
Moreover, as described in \autoref{sec:alignment}, the deflector can be used to create strongly
aligned and oriented samples, which allow for the investigation of steric effects in these chemical
reactions~\cite{Brooks:Science193:11, Miranda:NatPhys7:502}.

\subsection{Matrix deposition experiments}
\label{sec:outlook:matrix}
As a further application, one could envision to exploit the deflector for the deposition of selected
nuclear-spin isomers, structural isomers, enantiomers, or even specific molecular clusters in a
cryogenic rare gas matrix~\cite{Whittle:JCP22:1943, Bondybey:ChemRev96:2113}. Due to the extended
observation periods and the low temperatures in these matrices, such low-temperature storage would
allow for precision molecular spectroscopy as well as the investigation of slow dynamic processes,
similar to studies of nuclear-spin relaxation, for instance, of water~\cite{Turgeon:PRA86:062710,
   Sliter:JPCA115:9682}. These experiments could be extended to investigations of structural
isomerisation reactions or the bond-breaking and forming in specific clusters, providing novel
insight into these basic yet complex chemical transformations.

\section{Conclusions}
\label{sec:conclusions}
Inhomogeneous electric fields provide means to control the motion of even complex molecules. Modern
technologies and insight into these approaches, such as extreme supersonic expansions, strong,
possibly switched, electric fields and a detailed quantitative understanding of the interaction of
molecules with these strong fields has enabled the realisation of novel control experiments.
Exploiting appropriately shaped and timed fields, these approaches allow for the separation of
molecules based on their effective dipole moment, \ie, their internal state and, therefore, the
creation of gas-phase samples of molecules in selected quantum states, isomers, and cluster sizes.
We have described experiments that have purified beams of individual quantum states, isomers, and
sizes, as well as a number of first applications of these well-defined samples in chemical reaction
studies, further control experiments, and for ultrafast molecular imaging. In order to facilitate
the application of the electric deflector in further laboratories and experiments, we provide an
educational script to simulate the trajectories of molecules through an electric deflector
apparatus. Further applications have been detailed, which we consider feasible with the current
state of the art. They range from improved selectivity and the selection of considerably larger and
more complex molecules, over studies of fundamental effects and symmetries, to opportunities for
disentangling the chemical reaction kinetics and dynamics of large and complex molecules.

Overall, the development of a detailed, microscopic, reductionist understanding of the
structure-function relationship in the molecular sciences relies on the ability to produce clean
samples of pure ``molecules'', in the gas phase. Pushing these experiments into new regimes relies
on further technological advances. Especially for the far-reaching goal of controlling very large
biological ``molecules'', like proteins or viruses, novel concepts to control these particles are
required~\cite{Eckerskorn:OE21:30492, COMOTION:website}. Major advances in the field of time and
structure resolving imaging methods~\cite{Barty:ARPC64:415} will also be necessary. We believe that
the experiments described in this paper are an important step forward, but at the same time we
realise that ``there is plenty of room at the top'' to bridge small-molecule physical chemistry with
large-scale chemistry and (structural) biology.

\begin{acknowledgements}
   The experiments described in this article in which we have been involved are the result of more
   than ten years of research by a large group of people. The efforts started at the Department of
   Molecular Physics (headed by Gerard Meijer) of the Fritz Haber Institute of the Max Planck
   Society in 2003. In 2010, my group moved to the Center for Free-Electron Laser Science at DESY
   and the University of Hamburg. We are greatly indebted to the technical, administrative, and
   scientific staff at these institutions, and we thank all of the students, postdocs, senior
   scientists, and our external collaborators that have been involved in this work, and without whom
   these experiments would not have been possible.

   We wish to thank Frank Filsinger, Gerard Meijer, Jonas L.\ Hansen, Jochen Maurer, Jens H.\
   Nielsen, Lotte Holmegaard, and Henrik Stapelfeldt for help with the measurements presented in
   \autoref{fig:applications:3AP} and Joss Wiese for help with the measurements presented in
   \autoref{fig:applications:clusters}.

   Besides DESY, this work has been supported by the Helmholtz Association ``Initiative and
   Networking Fund'', the Max Planck Society, the \emph{Deutsche Forschungsgemeinschaft} (DFG)
   through the priority program ``SPP 1116: Interactions in ultracold atomic and molecular gases''
   (KU1527/1) and the excellence cluster ``The Hamburg Center for Ultrafast Imaging -- Structure,
   Dynamics and Control of Matter at the Atomic Scale'' (CUI, DFG-EXC1074), and the European
   Research Council through the Consolidator Grant 614507-COMOTION.
\end{acknowledgements}

\label{lastpage}
\small
\setlength{\bibsep}{2pt}
\bibliography{string,cmi}

\begin{thebibliography}{180}
\providecommand{\url}[1]{\texttt{#1}}
\providecommand{\urlprefix}{URL }
\markboth{Y.\,P.~Chang, D.~Horke, S.~Trippel, and J.~Küpper}{International
  Reviews in Physical Chemistry}

\bibitem{Kraitchman:AJP21:17}
J. Kraitchman,  Am.\ J.\ Phys.  \textbf{21}, 17 (1953).

\bibitem{Pratt:ARPC49:481}
D.~W. Pratt,  Annual Review of Physical Chemistry  \textbf{49} (1), 481--530
  (1998).
  $<$\url{http://physchem.annualreviews.org/cgi/content/abstract/49/1/481}$>$.

\bibitem{Wuethrich:ACIE42:3340}
K. W{\"u}thrich,  Angew.\ Chem.\ Int.\ Ed.  \textbf{42} (29), 3340--3363
  (2003).

\bibitem{Shipman:NewTechMW:2011}
S.~T. Shipman and B.~H. Pate, in \emph{Handbook of High-resolution
  Spectroscopy}, edited by M. Quack and F. Merkt  (John Wiley {\&} Sons, Ltd,
  Chichester, UK, 2011), Chap.~36, pp. 801--827.

\bibitem{Perez:Science336:897}
C. P\'{e}rez, M.~T. Muckle, D.~P. Zaleski, N.~a. Seifert, B. Temelso, G.~C.
  Shields, Z. Kisiel, and B.~H. Pate,  Science  \textbf{336} (6083), 897--901
  (2012).  $<$\url{http://www.ncbi.nlm.nih.gov/pubmed/22605772}$>$.

\bibitem{Barty:ARPC64:415}
A. Barty, J. K{\"u}pper, and H.~N. Chapman,  Annu.\ Rev.\ Phys.\ Chem.
  \textbf{64} (1), 415--435 (2013).
  $<$\url{http://dx.doi.org/10.1146/annurev-physchem-032511-143708}$>$.

\bibitem{Miller:Science343:1108}
R.~J.~D. Miller,  Science  \textbf{343} (6175), 1108--1116 (2014).

\bibitem{Ratzer:CP283:153}
C. Ratzer, J. K{\"u}pper, D. Spangenberg, and M. Schmitt,  Chem.\ Phys.
  \textbf{283} (1), 153--169 (2002).
  $<$\url{http://dx.doi.org/10.1016/S0301-0104(02)00591-8}$>$.

\bibitem{deLeeuw:CPL7:288}
F.~H. de~Leeuw and A. Dymanus,  Chem.\ Phys.\ Lett.  \textbf{7} (2), 288--292
  (1970).

\bibitem{Okruss:JCP110:10393}
M. Okruss, R. M{\"u}ller, and A. Hese,  J.\ Chem.\ Phys.  \textbf{110},
  10393--10402 (1999).

\bibitem{Royer:ChemRev106:1769}
C.~A. Royer,  Chem.\ Rev.  \textbf{106} (5), 1769–1784 (2006).

\bibitem{Mayes:JACS136:1008}
H.~B. Mayes, L.~J. Broadbelt, and G.~T. Beckham,  J.\ Am.\ Chem.\ Soc.
  \textbf{136} (3), 1008--1022 (2014).

\bibitem{Baldwin:NatChemBiol5:808}
A.~J. Baldwin and L.~E. Kay,  Nat. Chem. Biol.  \textbf{5} (11), 808--814
  (2009).

\bibitem{Blackmond:CSHPB2:a002147}
D.~G. Blackmond,  Cold Spring Harb.\ Perspect.\ Biol.  \textbf{2} (5), a002147
  (2010).

\bibitem{Pross:Life}
A. Pross, \emph{What is Life?: How chemistry becomes biology}   (, , 2012).

\bibitem{Suenram:JACS102:7180}
R.~D. Suenram and F.~J. Lovas,  J.\ Am.\ Chem.\ Soc.  \textbf{102}, 7180--7184
  (1980).

\bibitem{Rizzo:JCP83:4819}
T.~R. Rizzo, Y.~D. Park, L. Peteanu, and D.~H. Levy,  J.\ Chem.\ Phys.
  \textbf{83}, 4819--4820 (1985).

\bibitem{Scoles:MolBeam:1and2}
G. Scoles, editor, \emph{Atomic and molecular beam methods}, Vol. 1 \& 2
  (Oxford University Press, New York, NY, USA, 1988 \& 1992).

\bibitem{Fenn:ARPC47:1}
J.~B. Fenn,  Annu.\ Rev.\ Phys.\ Chem.  \textbf{47}, 1--41 (1996).

\bibitem{Erlekam:PCCP9:3786}
U. Erlekam, M. Frankowski, G. von Helden, and G. Meijer,  Phys.\ Chem.\ Chem.\
  Phys.  \textbf{9}, 3786--3789 (2007).

\bibitem{Meerakker:PRL95:013003}
S.~Y.~T. van~de Meerakker, N. Vanhaecke, M.~P.~J. van~der Loo, G.~C.
  Groenenboom, and G. Meijer,  Phys.\ Rev.\ Lett.  \textbf{95} (1), 013003
  (2005).  $<$\url{http://link.aps.org/abstract/PRL/v95/e013003}$>$.

\bibitem{Kuepper:FD142:155}
J. K{\"u}pper, F. Filsinger, and G. Meijer,  Faraday Disc.  \textbf{142},
  155--173 (2009).

\bibitem{Meerakker:NatPhys4:595}
S.~Y.~T. van~de Meerakker, H.~L. Bethlem, and G. Meijer,  Nat. Phys.
  \textbf{4} (8), 595 (2008).  $<$\url{http://dx.doi.org/10.1038/nphys1031}$>$.

\bibitem{Schnell:ACIE48:6010}
M. Schnell and G. Meijer,  Angew.\ Chem.\ Int.\ Ed.  \textbf{48} (33),
  6010--6031 (2009).  $<$\url{http://dx.doi.org/10.1002/anie.200805503}$>$.

\bibitem{Meerakker:CR112:2012}
S.~Y.~T. van~de Meerakker, H.~L. Bethlem, N. Vanhaecke, and G. Meijer,  Chem.\
  Rev.  \textbf{112} (9), 4828--4878 (2012).
  $<$\url{http://pubs.acs.org/doi/abs/10.1021/cr200349r}$>$.

\bibitem{Bethlem:Nature406:491}
H.~L. Bethlem, G. Berden, F.~M.~H. Crompvoets, R.~T. Jongma, A.~J.~A. van Roij,
  and G. Meijer,  Nature  \textbf{406}, 491--494 (2000).
  $<$\url{http://dx.doi.org/10.1038/35020030}$>$.

\bibitem{Loesch:JCP93:4779}
H.~J. Loesch and A. Remscheid,  J.\ Chem.\ Phys.  \textbf{93}, 4779 (1990).

\bibitem{Friedrich:PRL74:4623}
B. Friedrich and D. Herschbach,  Phys.\ Rev.\ Lett.  \textbf{74} (23),
  4623--4626 (1995).

\bibitem{Stapelfeldt:RMP75:543}
H. Stapelfeldt and T. Seideman,  Rev.\ Mod.\ Phys.  \textbf{75} (2), 543--557
  (2003).  $<$\url{http://link.aps.org/abstract/RMP/v75/p543}$>$.

\bibitem{Holmegaard:PRL102:023001}
L. Holmegaard, J.~H. Nielsen, I. Nevo, H. Stapelfeldt, F. Filsinger, J.
  K{\"u}pper, and G. Meijer,  Phys.\ Rev.\ Lett.  \textbf{102}, 023001 (2009).
  $<$\url{http://dx.doi.org/10.1103/PhysRevLett.102.023001}$>$.

\bibitem{Kallmann:ZP6:352}
H. Kallmann and F. Reiche,  Z.\ Phys.  \textbf{6}, 352--375 (1921).
  $<$\url{http://dx.doi.org/10.1007/BF01327996}$>$.

\bibitem{Gerlach:ZP9:349}
W. Gerlach and O. Stern,  Z.\ Phys.  \textbf{9}, 349--352 (1922).
  $<$\url{http://dx.doi.org/10.1007/BF01326983}$>$.

\bibitem{Wrede:ZP44:261}
E. Wrede,  Z.\ Phys.  \textbf{44} (4-5), 261--268 (1927).
  $<$\url{http://dx.doi.org/10.1007/BF01391193}$>$.

\bibitem{Vanhaecke:PRA75:031402}
N. Vanhaecke, U. Meier, M. Andrist, B.~H. Meier, and F. Merkt,  Phys.\ Rev.\ A
  \textbf{75} (3), 031402(R) (2007).

\bibitem{Narevicius:CR112:4879}
E. Narevicius and M.~G. Raizen,  Chem.\ Rev.  \textbf{112} (9), 4879--89
  (2012).  $<$\url{http://www.ncbi.nlm.nih.gov/pubmed/22827566}$>$.

\bibitem{Stern:ZP39:751}
O. Stern,  Z.\ Phys.  \textbf{39} (10-11), 751--763 (1926).
  $<$\url{http://dx.doi.org/10.1007/BF01451746}$>$.

\bibitem{Rabi:PR55:526}
I.~I. Rabi, S. Millman, P. Kusch, and J.~R. Zacharias,  Phys.\ Rev.
  \textbf{55} (6), 526--535 (1939).
  $<$\url{http://journals.aps.org/pr/abstract/10.1103/PhysRev.55.526}$>$.

\bibitem{Gordon:PR95:282}
J.~P. Gordon, H.~J. Zeiger, and C.~H. Townes,  Phys.\ Rev.  \textbf{95},
  282--284 (1954).

\bibitem{Bennewitz:ZP141:6}
H.~G. Bennewitz, W. Paul, and C. Schlier,  Z.\ Phys.  \textbf{141}, 6 (1955).

\bibitem{Gordon:PR99:1253}
J.~P. Gordon,  Phys.\ Rev.  \textbf{99}, 1253--1263 (1955).

\bibitem{Gordon:PR99:1264}
J.~P. Gordon, H.~J. Zeiger, and C.~H. Townes,  Phys.\ Rev.  \textbf{99},
  1264--1274 (1955).

\bibitem{Veldhoven:EPJD31:337}
J. van Veldhoven, J. K{\"u}pper, H.~L. Bethlem, B. Sartakov, A.~J.~A. van Roij,
  and G. Meijer,  Eur.\ Phys.\ J.\ D  \textbf{31} (2), 337--349 (2004).
  $<$\url{http://dx.doi.org/10.1140/epjd/e2004-00160-9}$>$.

\bibitem{Hillenkamp:JCP118:8699}
M. Hillenkamp, S. Keinan, and U. Even,  J.\ Chem.\ Phys.  \textbf{118} (19),
  8699--8705 (2003).  $<$\url{http://link.aip.org/link/?JCP/118/8699/1}$>$.

\bibitem{Jin:CR112:4801}
D.~S. Jin and J. Ye,  Chem.\ Rev.  \textbf{112} (9), 4801--4802 (2012).

\bibitem{Auerbach:JCP45:2160}
D. Auerbach, E.~E.~A. Bromberg, and L. Wharton,  J.\ Chem.\ Phys.  \textbf{45},
  2160 (1966).

\bibitem{Guenther:ZPCNF80:155}
F. G\"{u}nther and K. Sch{\"u}gerl,  Z.\ Phys.\ Chem.  \textbf{NF\,80}, 155
  (1972).

\bibitem{Filsinger:PRL100:133003}
F. Filsinger, U. Erlekam, G. von Helden, J. K{\"u}pper, and G. Meijer,  Phys.\
  Rev.\ Lett.  \textbf{100}, 133003 (2008).
  $<$\url{http://dx.doi.org/10.1103/PhysRevLett.100.133003}$>$.

\bibitem{Filsinger:PRA82:052513}
F. Filsinger, S. Putzke, H. Haak, G. Meijer, and J. K{\"u}pper,  Phys.\ Rev.\ A
   \textbf{82}, 052513 (2010).

\bibitem{Putzke:PCCP13:18962}
S. Putzke, F. Filsinger, H. Haak, J. K{\"u}pper, and G. Meijer,  Phys.\ Chem.\
  Chem.\ Phys.  \textbf{13}, 18962 (2011).

\bibitem{Bethlem:PRL88:133003}
H.~L. Bethlem, A.~J.~A. van Roij, R.~T. Jongma, and G. Meijer,  Phys.\ Rev.\
  Lett.  \textbf{88} (13), 133003 (2002).
  $<$\url{http://doi.dx.org/10.1103/PhysRevLett.88.133003}$>$.

\bibitem{Tarbutt:PRL92:173002}
M.~R. Tarbutt, H.~L. Bethlem, J.~J. Hudson, V.~L. Ryabov, V.~A. Ryzhov, B.~E.
  Sauer, G. Meijer, and E.~A. Hinds,  Phys.\ Rev.\ Lett.  \textbf{92} (17),
  173002 (2004).  $<$\url{http://dx.doi.org/10.1103/PhysRevLett.92.173002}$>$.

\bibitem{Bethlem:JPB39:R263}
H.~L. Bethlem, M.~R. Tarbutt, J. K{\"u}pper, D. Carty, K. Wohlfart, E.~A.
  Hinds, and G. Meijer,  J.\ Phys.\ B  \textbf{39} (16), R263--R291 (2006).
  $<$\url{http://dx.doi.org/10.1088/0953-4075/39/16/R01}$>$.

\bibitem{Wohlfart:PRA78:033421}
K. Wohlfart, F. Filsinger, F. Gr{\"a}tz, J. K{\"u}pper, and G. Meijer,  Phys.\
  Rev.\ A  \textbf{78} (3), 033421 (2008).

\bibitem{Wohlfart:PRA77:031404}
K. Wohlfart, F. Gr{\"a}tz, F. Filsinger, H. Haak, G. Meijer, and J. K{\"u}pper,
   Phys.\ Rev.\ A  \textbf{77}, 031404(R) (2008).
  $<$\url{http://dx.doi.org/10.1103/PhysRevA.77.031404}$>$.

\bibitem{Filsinger:PCCP13:2076}
F. Filsinger, G. Meijer, H. Stapelfeldt, H. Chapman, and J. K{\"u}pper,  Phys.\
  Chem.\ Chem.\ Phys.  \textbf{13} (6), 2076--2087 (2011).

\bibitem{Odashima:PRL104:253001}
H. Odashima, S. Merz, K. Enomoto, M. Schnell, and G. Meijer,  Phys.\ Rev.\
  Lett.  \textbf{104}, 253001 (2010).

\bibitem{Stapelfeldt:PRL79:2787}
H. Stapelfeldt, H. Sakai, E. Constant, and P.~B. Corkum,  Phys.\ Rev.\ Lett.
  \textbf{79}, 2787--2790 (1997).
  $<$\url{http://dx.doi.org/10.1103/PhysRevLett.79.2787}$>$.

\bibitem{Zhao:PRL85:2705}
B.~S. Zhao, H.~S. Chung, K. Cho, S.~H. Lee, S. Hwang, J. Yu, Y.~H. Ahn, J.~Y.
  Sohn, D.~S. Kim, W.~K. Kang, and D.~S. Chung,  Phys.\ Rev.\ Lett.
  \textbf{85} (13), 2705--2708 (2000).

\bibitem{Fulton:NatPhys2:465}
R. Fulton, A.~I. Bishop, M.~N. Shneider, and P.~F. Barker,  Nat. Phys.
  \textbf{2}, 465--468 (2006).

\bibitem{Ashkin:PRL24:156}
A. Ashkin,  Phys.\ Rev.\ Lett.  \textbf{24} (4), 156--159 (1970).
  $<$\url{http://link.aps.org/doi/10.1103/PhysRevLett.24.156}$>$.

\bibitem{Lebedew:AnaPhys311:433}
P. Lebedew,  Ann.\ Phys.  \textbf{311} (11), 433--458 (1901).
  $<$\url{http://doi.wiley.com/10.1002/andp.19013111102}$>$.

\bibitem{Ehrenhaft:AnaPhys361:81}
F. Ehrenhaft,  Ann.\ Phys.  \textbf{361} (10), 81--132 (1918).
  $<$\url{http://doi.wiley.com/10.1002/andp.19183611002}$>$.

\bibitem{Eckerskorn:OE21:30492}
N. Eckerskorn, L. Li, R.~A. Kirian, J. K{\"u}pper, D.~P. DePonte, W.
  Krolikowski, W.~M. Lee, H.~N. Chapman, and A.~V. Rode,  Opt.\ Exp.
  \textbf{21} (25), 30492--30499 (2013).

\bibitem{COMOTION:website}
Controlling the Motion of Complex Molecules and Particles (COMOTION)  2015.
  $<$\url{http://www.comotion.info}$>$.

\bibitem{Rizzo:JCP84:2534}
T.~R. Rizzo, Y.~D. Park, L.~A. Peteanu, and D.~H. Levy,  J.\ Chem.\ Phys.
  \textbf{84}, 2534--2541 (1986).

\bibitem{Vries:ARPC58:585}
M.~S. de~Vries and P. Hobza,  Annu.\ Rev.\ Phys.\ Chem.  \textbf{58}, 585--612
  (2007).

\bibitem{Neill:PCCP13:7253}
J.~L. Neill, K.~O. Douglass, B.~H. Pate, and D.~W. Pratt,  Phys.\ Chem.\ Chem.\
  Phys.  \textbf{13} (16), 7253--7262 (2011).

\bibitem{Watson:Nature171:737}
J. Watson and F. Crick,  Nature  \textbf{171} (4356), 737--738 (1953).
  $<$\url{http://dx.doi.org/10.1038/171737a0}$>$.

\bibitem{Hedberg:Science254:410}
K. Hedberg, L. Hedberg, D.~S. Bethune, C.~A. Brown, H.~C. Dorn, R.~D. Johnson,
  and M. de~Vries,  Science  \textbf{254} (5030), 410--412 (1991).
  $<$\url{http://www.sciencemag.org/cgi/content/abstract/254/5030/410}$>$.

\bibitem{Williamson:Nature386:159}
J.~C. Williamson, J.~M. Cao, H. Ihee, H. Frey, and A.~H. Zewail,  Nature
  \textbf{386} (6621), 159--162 (1997).

\bibitem{Sciaini:RPP74:096101}
G. Sciaini and R.~J.~D. Miller,  Rep.\ Prog.\ Phys.  \textbf{74} (9), 096101
  (2011).  $<$\url{http://iopscience.iop.org/0034-4885/74/9/096101/}$>$.

\bibitem{Neutze:Nature406:752}
R. Neutze, R. Wouts, D. van~der Spoel, E. Weckert, and J. Hajdu,  Nature
  \textbf{406} (6797), 752--757 (2000).
  $<$\url{http://dx.doi.org/10.1038/35021099}$>$.

\bibitem{Hensley:PRL109:133202}
C.~J. Hensley, J. Yang, and M. Centurion,  Phys.\ Rev.\ Lett.  \textbf{109},
  133202 (2012).

\bibitem{Kuepper:PRL112:083002}
J. K{\"u}pper, S. Stern, L. Holmegaard, F. Filsinger, A. Rouz\'{e}e, A.
  Rudenko, P. Johnsson, A.~V. Martin, M. Adolph, A. Aquila, S. Bajt, A. Barty,
  {\em{et~al.}},  Phys.\ Rev.\ Lett.  \textbf{112}, 083002 (2014).

\bibitem{Reid:MolPhys3:131}
K.~L. Reid,  Molecular Physics  \textbf{3}, 131--147 (2012).

\bibitem{Bisgaard:Science323:1464}
C.~Z. Bisgaard, O.~J. Clarkin, G. Wu, A.~M.~D. Lee, O. Ge{\ss}ner, C.~C.
  Hayden, and A. Stolow,  Science  \textbf{323} (5920), 1464--1468 (2009).

\bibitem{Holmegaard:NatPhys6:428}
L. Holmegaard, J.~L. Hansen, L. Kalh{\o}j, S.~L. Kragh, H. Stapelfeldt, F.
  Filsinger, J. K{\"u}pper, G. Meijer, D. Dimitrovski, M. Abu-samha, C.~P.~J.
  Martiny, and L.~B. Madsen,  Nat. Phys.  \textbf{6}, 428 (2010).

\bibitem{Akagi:Science325:1364}
H. Akagi, T. Otobe, A. Staudte, A. Shiner, F. Turner, R. D{\"o}rner, D.
  Villeneuve, and P. Corkum,  Science  \textbf{325} (5946), 1364--1367 (2009).

\bibitem{Hansen:PRL106:073001}
J. Hansen, H. Stapelfeldt, D. Dimitrovski, M. Abu-Samha, C. Martiny, and L.
  Madsen,  Phys.\ Rev.\ Lett.  \textbf{106} (7), 073001 (2011).
  $<$\url{http://prl.aps.org/abstract/PRL/v106/i7/e073001}$>$.

\bibitem{Maurer:PRL109:123001}
J. Maurer, D. Dimitrovski, L. Christensen, L.~B. Madsen, and H. Stapelfeldt,
  Phys.\ Rev.\ Lett.  \textbf{109}, 123001 (2012).
  $<$\url{http://link.aps.org/doi/10.1103/PhysRevLett.109.123001}$>$.

\bibitem{Zuo:CPL159:313}
T. Zuo, A.~D. Bandrauk, and P.~B. Corkum,  Chem.\ Phys.\ Lett.  \textbf{259}
  (3-4), 313--320 (1996).

\bibitem{Blaga:Nature483:194}
C.~I. Blaga, J. Xu, A.~D. DiChiara, E. Sistrunk, K. Zhang, P. Agostini, T.~A.
  Miller, L.~F. DiMauro, and C.~D. Lin,  Nature  \textbf{483} (7388), 194--197
  (2012).

\bibitem{Itatani:Nature432:867}
J. Itatani, J. Levesque, D. Zeidler, H. Niikura, H. P\'{e}pin, J.~C. Kieffer,
  P.~B. Corkum, and D.~M. Villeneuve,  Nature  \textbf{432}, 867--871 (2004).

\bibitem{Woerner:Nature466:604}
H.~J. W{\"o}rner, J.~B. Bertrand, D.~V. Kartashov, P.~B. Corkum, and D.~M.
  Villeneuve,  Nature  \textbf{466} (7306), 604--607 (2010).

\bibitem{Landers:PRL87:013002}
A. Landers, T. Weber, I. Ali, A. Cassimi, M. Hattass, O. Jagutzki, A. Nauert,
  T. Osipov, A. Staudte, M.~H. Prior, H. Schmidt-B{\"o}cking, C.~L. Cocke,
  {\em{et~al.}},  Phys.\ Rev.\ Lett.  \textbf{87} (1), 013002 (2001).
  $<$\url{http://prola.aps.org/abstract/PRL/v87/i1/e013002}$>$.

\bibitem{Krasniqi:PRA81:033411}
F. Krasniqi, B. Najjari, L. Str\"{u}der, D. Rolles, A. Voitkiv, and J. Ullrich,
   Phys.\ Rev.\ A  \textbf{81}, 033411 (2010).

\bibitem{Boll:PRA88:061402}
R. Boll, D. Anielski, C. Bostedt, J.~D. Bozek, L. Christensen, R. Coffee, S.
  De, P. Decleva, S.~W. Epp, B. Erk, L. Foucar, F. Krasniqi, {\em{et~al.}},
  Phys.\ Rev.\ A  \textbf{88}, 061402(R) (2013).

\bibitem{Lee:Science236:4803}
Y.~T. Lee,  Science  \textbf{236} (4803), 793--798 (1987).

\bibitem{Zewail:JPCA104:5660}
A.~H. Zewail,  J.\ Phys.\ Chem.\ A  \textbf{104} (24), 5660--5694 (2000).

\bibitem{Warren:Science259:1581}
W.~S. Warren, H. Rabitz, and M. Dahleh,  Science  \textbf{259}, 1581--1589
  (1993).  $<$\url{http://www.jstor.org/stable/2880660}$>$.

\bibitem{Assion:Science282:919}
A. Assion, T. Baumert, M. Bergt, T. Brixner, B. Kiefer, V. Seyfried, M.
  Strehle, and G. Gerber,  Science  \textbf{282}, 919--922 (1998).
  $<$\url{http://dx.doi.org/10.1126/science.282.5390.919}$>$.

\bibitem{Wang:Science342:1499}
T. Wang, J. Chen, T. Yang, C. Xiao, Z. Sun, L. Huang, D. Dai, X. Yang, and
  D.~H. Zhang,  Science  \textbf{342} (6165), 1499--1502 (2013).

\bibitem{Vogels:PRL113:263202}
S.~N. Vogels, J. Onvlee, A. von Zastrow, G.~C. Groenenboom, A. van~der Avoird,
  and S.~Y.~T. Meerakker,  Phys.\ Rev.\ Lett.  \textbf{113} (26), 263202
  (2014).

\bibitem{Ihee:Science291:458}
H. Ihee, V. Lobastov, U. Gomez, B. Goodson, R. Srinivasan, C. Ruan, and A.~H.
  Zewail,  Science  \textbf{291} (5503), 458--462 (2001).

\bibitem{Gordy:MWMolSpec}
W. Gordy and R.~L. Cook, \emph{Microwave Molecular Spectra}, 3rd ed.   (John
  Wiley \& Sons, New York, NY, USA, 1984).

\bibitem{Chang:CPC185:339}
Y.~P. Chang, F. Filsinger, B. Sartakov, and J. K{\"u}pper,  Comp.\ Phys.\ Comm.
   \textbf{185}, 339--49 (2014).
  $<$\url{http://www.sciencedirect.com/science/article/pii/S0010465513003019}$>$.

\bibitem{Wohlfart:JMolSpec247:119}
K. Wohlfart, M. Schnell, J.~U. Grabow, and J. K{\"u}pper,  J.\ Mol.\ Spec.
  \textbf{247} (1), 119--121 (2008).
  $<$\url{http://dx.doi.org/10.1016/j.jms.2007.10.006}$>$.

\bibitem{AbdElRahim:JPCA109:8507}
M. {Abd El Rahim}, R. Antoine, M. Broyer, D. Rayane, and P. Dugourd,  J.\
  Phys.\ Chem.\ A  \textbf{109} (38), 8507--14 (2005).
  $<$\url{http://www.ncbi.nlm.nih.gov/pubmed/16834247}$>$.

\bibitem{Ramsey:MolBeam:1956}
N.~F. Ramsey, \emph{Molecular Beams} The International Series of Monographs on
  Physics  (Oxford University Press, London, GB, 1956 \unskip; reprinted in
  \emph{Oxford Classic Texts in the Physical Sciences} (2005)), reprinted in
  \emph{Oxford Classic Texts in the Physical Sciences} (2005).

\bibitem{Filsinger:JCP131:064309}
F. Filsinger, J. K{\"u}pper, G. Meijer, L. Holmegaard, J.~H. Nielsen, I. Nevo,
  J.~L. Hansen, and H. Stapelfeldt,  J.\ Chem.\ Phys.  \textbf{131}, 064309
  (2009).
  $<$\url{http://scitation.aip.org/content/aip/journal/jcp/131/6/10.1063/1.3194287}$>$.

\bibitem{Kalnins:RSI73:2557}
J.~G. Kalnins, G. Lambertson, and H. Gould,  Rev.\ Sci.\ Instrum.  \textbf{73},
  2557--2565 (2002).
  $<$\url{http://scitation.aip.org/content/aip/journal/rsi/73/7/10.1063/1.1485778}$>$.

\bibitem{DeNijs:PCCP13:19052}
A.~J. de~Nijs and H.~L. Bethlem,  Phys.\ Chem.\ Chem.\ Phys.  \textbf{13} (42),
  19052--8 (2011).  $<$\url{http://www.ncbi.nlm.nih.gov/pubmed/21979152}$>$.

\bibitem{Miller:AAMOP25:37}
T.~M. Miller and B. Bederson,  Adv.\ Atom.\ Mol.\ Opt.\ Phys.  \textbf{25},
  37--60 (1988).

\bibitem{Schaefer:PRL76:471}
R. Sch{\"a}fer, S. Schlecht, J. Woenckhaus, and J. Becker,  Phys.\ Rev.\ Lett.
  \textbf{76} (3), 471--474 (1996).
  $<$\url{http://prl.aps.org/abstract/PRL/v76/i3/p471_1}$>$.

\bibitem{deHeer:arXiv0901:4810}
W.~A. de~Heer and V.~V. Kresin,  arXiv  p. 0901.4810 (2009).
  $<$\url{http://de.arxiv.org/abs/0901.4810}$>$, Article prepared for the
  Handbook of Nanophysics, ed.\ by Klaus D.\ Sattler, \emph{to be published} by
  Taylor\&Francis/CRC Press.

\bibitem{Trippel:MP111:1738}
S. Trippel, T. Mullins, N.~L.~M. M{\"u}ller, J.~S. Kienitz, K.
  D{\l}ugo{\l}\k{e}cki, and J. K{\"u}pper,  Mol.\ Phys.  \textbf{111}, 1738
  (2013).

\bibitem{deNijs:JMolSpec300:79}
A.~J. de~Nijs, W. Ubachs, and H.~L. Bethlem,  J.\ Mol.\ Spec.  \textbf{300},
  79--85 (2014).

\bibitem{Nishimura:RSI74:3271}
H. Nishimura, G. Lambertson, J.~G. Kalnins, and H. Gould,  Rev.\ Sci.\ Instrum.
   \textbf{74} (7), 3271--2378 (2003).
  $<$\url{http://doi.dx.org/10.1063/1.1578159}$>$.

\bibitem{Putzke:thesis:2012}
S. Putzke, Alternating-gradient focusing of large neutral molecules,
  Dissertation, Freie Universit{\"a}t, {B}erlin 2012.

\bibitem{Trippel:PRA86:033202}
S. Trippel, Y.~P. Chang, S. Stern, T. Mullins, L. Holmegaard, and J.
  K{\"u}pper,  Phys.\ Rev.\ A  \textbf{86}, 033202 (2012).
  $<$\url{http://pra.aps.org/abstract/PRA/v86/i3/e033202}$>$.

\bibitem{Luria:JPCA115:7362}
K. Luria, W. Christen, and U. Even,  J.\ Phys.\ Chem.\ A  \textbf{115} (25),
  7362--7367 (2011).

\bibitem{Escribano:PRA62:023407}
R. Escribano, B. Mate, F. Ortigoso, and J. Ortigoso,  Phys.\ Rev.\ A
  \textbf{62} (2), 023407 (2000).

\bibitem{Wall:PRA81:033414}
T.~E. Wall, S.~K. Tokunaga, E.~a. Hinds, and M.~R. Tarbutt,  Phys.\ Rev.\ A
  \textbf{81} (3), 033414 (2010).
  $<$\url{http://link.aps.org/doi/10.1103/PhysRevA.81.033414}$>$.

\bibitem{Kirste:PRA79:051401}
M. Kirste, B.~G. Sartakov, M. Schnell, and G. Meijer,  Phys.\ Rev.\ A
  \textbf{79}, 051401(R) (2009).
  $<$\url{http://dx.doi.org/10.1103/PhysRevA.79.051401}$>$.

\bibitem{Veldhoven:PRA66:032501}
J. van Veldhoven, R.~T. Jongma, B. Sartakov, W.~A. Bongers, and G. Meij\-er,
  Phys.\ Rev.\ A  \textbf{66} (3), 32501 (2002).

\bibitem{Filsinger:thesis:2010}
F. Filsinger, Manipulation of large neutral molecules with electric fields,
  Ph.D. thesis, Radboud Universiteit, Nijmegen, The Netherlands 2010.

\bibitem{Nielsen:PCCP13:18971}
J.~H. Nielsen, P. Simesen, C.~Z. Bisgaard, H. Stapelfeldt, F. Filsinger, B.
  Friedrich, G. Meijer, and J. K{\"u}pper,  Phys.\ Chem.\ Chem.\ Phys.
  \textbf{13}, 18971--18975 (2011).

\bibitem{Trippel:PRA89:051401R}
S. Trippel, T. Mullins, N.~L.~M. M{\"u}ller, J.~S. Kienitz, J.~J. Omiste, H.
  Stapelfeldt, R. Gonz{\'a}lez-F{\'e}rez, and J. K{\"u}pper,  Phys.\ Rev.\ A
  \textbf{89}, 051401(R) (2014).

\bibitem{Horke:ACIE53:11965}
D.~A. Horke, Y.~P. Chang, K. D{\l}ugo{\l}\k{e}cki, and J. K{\"u}pper,  Angew.\
  Chem.\ Int.\ Ed.  \textbf{53}, 11965--11968 (2014).
  $<$\url{http://onlinelibrary.wiley.com/doi/10.1002/anie.201405986/abstract}$>$.

\bibitem{Filsinger:ACIE48:6900}
F. Filsinger, J. K{\"u}pper, G. Meijer, J.~L. Hansen, J. Maurer, J.~H. Nielsen,
  L. Holmegaard, and H. Stapelfeldt,  Angew.\ Chem.\ Int.\ Ed.  \textbf{48},
  6900--6902 (2009).
  $<$\url{http://onlinelibrary.wiley.com/doi/10.1002/anie.200902650/abstract}$>$.

\bibitem{Kierspel:CPL591:130}
T. Kierspel, D.~A. Horke, Y.~P. Chang, and J. K{\"u}pper,  Chem.\ Phys.\ Lett.
  \textbf{591}, 130--132 (2014).
  $<$\url{http://www.sciencedirect.com/science/article/pii/S0009261413014000}$>$.

\bibitem{Chang:Science342:98}
Y.~P. Chang, K. D{\l}ugo\l\k{e}cki, J. K{\"u}pper, D. R{\"o}sch, D. Wild, and
  S. Willitsch,  Science  \textbf{342} (6154), 98--101 (2013).
  $<$\url{http://dx.doi.org/10.1126/science.1242271}$>$.

\bibitem{Helden:Science267:1483}
G. von Helden, T. Wyttenbach, and M.~T. Bowers,  Science  \textbf{267},
  1483--1485 (1995).

\bibitem{Jarrold:PCCP9:1659}
M. Jarrold,  Phys.\ Chem.\ Chem.\ Phys.  \textbf{9}, 1659--1671 (2007).

\bibitem{Papadopoulos:FD150:243}
G. Papadopoulos, A. Svendsen, O.~V. Boyarkin, and T.~R. Rizzo,  Faraday Disc.
  \textbf{150}, 243--255 (2011).
  $<$\url{http://dx.doi.org/10.1039/C0FD00004C}$>$.

\bibitem{Kanu:JMS43:1}
A.~B. Kanu, P. Dwivedi, M. Tam, L. Matz, and H.~H. Hill,  J.\ Mass.\ Spectrom.
  \textbf{43} (1), 1--22 (2008).

\bibitem{Park:Nature415:306}
S.~T. Park, S.~K. Kim, and M.~S. Kim,  Nature  \textbf{415}, 306 (2002).
  $<$\url{http://www.nature.com/nature/journal/v415/n6869/abs/415306a.html}$>$.

\bibitem{Kim:Science315:1561}
M.~H. Kim, L. Shen, H. Tao, T.~J. Martinez, and A.~G. Suits,  Science
  \textbf{315}, 1561 (2007).
  $<$\url{http://www.sciencemag.org/content/315/5818/1561}$>$.

\bibitem{Dian:Science303:1169}
B.~C. Dian, J.~R. Clarkson, and T.~S. Zwier,  Science  \textbf{303} (5661),
  1169--1173 (2004).
  $<$\url{http://www.sciencemag.org/cgi/content/abstract/303/5661/1169}$>$.

\bibitem{Dian:Science296:2369}
B.~C. Dian, A. Longarte, and T.~S. Zwier,  Science  \textbf{296} (5577),
  2369--2373 (2002).
  $<$\url{http://www.sciencemag.org/cgi/content/abstract/296/5577/2369}$>$.

\bibitem{Wohlfart:thesis:2008}
K. Wohlfart, Alternating-gradient focusing and deceleration of large molecules,
  Dissertation, Free University, {B}erlin, Germany 2008.
  $<$\url{http://www.diss.fu-berlin.de/diss/receive/FUDISS_thesis_000000004336}$>$.

\bibitem{Horke:JoVE:e51137}
D.~A. Horke, S. Trippel, Y.~P. Chang, S. Stern, T. Mullins, T. Kierspel, and J.
  K{\"u}pper,  J.\ Vis.\ Exp.  p. e51137 (2014).
  $<$\url{http://www.jove.com/video/51137/spatial-separation-of-molecular-conformers-and-clusters}$>$.

\bibitem{Trippel:PRL114:103003}
S. Trippel, T. Mullins, N.~L.~M. M{\"u}ller, J.~S. Kienitz, R.
  Gonz{\'a}lez-F{\'e}rez, and J. K{\"u}pper,  Phys.\ Rev.\ Lett.  \textbf{114},
  103003 (2015).

\bibitem{Roesch:JCP140:124202}
D. R{\"o}sch, S. Willitsch, Y.~P. Chang, and J. K{\"u}pper,  J.\ Chem.\ Phys.
  \textbf{140} (12), 124202 (2014).
  $<$\url{http://dx.doi.org/10.1063/1.4869100}$>$.

\bibitem{Willitsch:IRPC31:175}
S. Willitsch,  Int.\ Rev.\ Phys.\ Chem.  \textbf{31}, 175 (2012).
  $<$\url{http://www.tandfonline.com/doi/abs/10.1080/0144235X.2012.667221}$>$.

\bibitem{Tong:PRL105:143001}
X. Tong, A.~H. Winney, and S. Willitsch,  Phys.\ Rev.\ Lett.  \textbf{105},
  143001 (2010).
  $<$\url{http://link.aps.org/doi/10.1103/PhysRevLett.105.143001}$>$.

\bibitem{Verlet:CSR37:505}
J.~R.~R. Verlet,  Chem.\ Soc.\ Rev.  \textbf{37}, 505--517 (2008).

\bibitem{Fujii:IRPC32:266}
A. Fujii and K. Mizuse,  Int.\ Rev.\ Phys.\ Chem.  \textbf{32} (2), 266--307
  (2013).

\bibitem{Buck:PRL52:109}
U. Buck and H. Meyer,  Phys.\ Rev.\ Lett.  \textbf{52} (2), 109--112 (1984).
  $<$\url{http://link.aps.org/doi/10.1103/PhysRevLett.52.109}$>$.

\bibitem{Goerke:ZPD19:137}
A. Goerke, M. Feser, H. Palm, C. Schulz, and I. Hertel,  Z.\ Phys.\ D
  \textbf{19} (4), 137--139 (1991).

\bibitem{Buck:CR100:3863}
U. Buck and F. Huisken,  Chem.\ Rev.  \textbf{100} (11), 3863--3890 (2000).
  $<$\url{http://dx.doi.org/10.1021/cr990054v}$>$.

\bibitem{Putzke:JCP137:104310}
S. Putzke, F. Filsinger, J. K{\"u}pper, and G. Meijer,  J.\ Chem.\ Phys.
  \textbf{137} (10), 104310 (2012).
  $<$\url{http://link.aip.org/link/?JCPSA6/137/104310/1}$>$.

\bibitem{Kang:JCP122:174301}
C. Kang, T.~M. Korter, and D.~W. Pratt,  J.\ Chem.\ Phys.  \textbf{122} (17),
  174301 (2005).  $<$\url{http://link.aip.org/link/?JCP/122/174301/1}$>$.

\bibitem{Bennewitz:ZP139:489}
H.~G. Bennewitz and W. Paul,  Z.\ Phys.  \textbf{139}, 489 (1954).
  $<$\url{http://dx.doi.org/10.1007/BF01374557}$>$.

\bibitem{Reuss:StateSelection}
J. Reuss, in \emph{Atomic and molecular beam methods}, edited by G. Scoles,
  Vol.~1, Chap.~11  (Oxford University Press, New York, NY, USA, 1988), pp.
  276--292.

\bibitem{Bethlem:PRL83:1558}
H.~L. Bethlem, G. Berden, and G. Meijer,  Phys.\ Rev.\ Lett.  \textbf{83},
  1558--1561 (1999).  $<$\url{http://link.aps.org/abstract/PRL/v83/p1558}$>$.

\bibitem{Brouard:CSR:2014}
M. Brouard, D.~H. Parker, and S.~Y.~T. van~de Meerakker,  Chem.\ Soc.\ Rev.
  (2014).  $<$\url{http://dx.doi.org/10.1039/c4cs00150h}$>$.

\bibitem{Bennewitz:ZP177:84}
H.~G. Bennewitz, K.~H. Kramer, J.~P. Toennies, and W. Paul,  Z.\ Phys.
  \textbf{177} (1), 84 (1964).

\bibitem{Brooks:JCP45:3449}
P.~R. Brooks and E.~M. Jones,  J.\ Chem.\ Phys.  \textbf{45} (9), 3449 (1966).

\bibitem{Beuhler:JACS88:5331}
R.~J. Beuhler, R.~B. Bernstein, and K.~H. Kramer,  J.\ Am.\ Chem.\ Soc.
  \textbf{88} (22), 5331 (1966).

\bibitem{Kumarappan:JCP125:194309}
V. Kumarappan, C.~Z. Bisgaard, S.~S. Viftrup, L. Holmegaard, and H.
  Stapelfeldt,  J.\ Chem.\ Phys.  \textbf{125} (19), 194309 (2006).

\bibitem{Nevo:PCCP11:9912}
I. Nevo, L. Holmegaard, J.~H. Nielsen, J.~L. Hansen, H. Stapelfeldt, F.
  Filsinger, G. Meijer, and J. K{\"u}pper,  Phys.\ Chem.\ Chem.\ Phys.
  \textbf{11}, 9912--9918 (2009).

\bibitem{Rolles:JPB47:124035}
D. Rolles, R. Boll, M. Adolph, A. Aquila, C. Bostedt, J. Bozek, H. Chapman, R.
  Coffee, N. Coppola, P. Decleva, T. Delmas, S. Epp, {\em{et~al.}},  J.\ Phys.\
  B  \textbf{47} (12), 124035 (2014).

\bibitem{Stern:FD171:393}
S. Stern, L. Holmegaard, F. Filsinger, A. Rouzee, A. Rudenko, P. Johnsson,
  A.~V. Martin, A. Barty, C. Bostedt, J. Bozek, R. Coffee, S. Epp,
  {\em{et~al.}},  Faraday Disc.  \textbf{171}, 393 (2014).

\bibitem{Calvo:BiophysJ95:18}
F. Calvo and P. Dugourd,  Biophys. J.  \textbf{95} (1), 18--32 (2008).

\bibitem{Abrikossov:thesis:2011}
A. Abrikossov, Computer simulation of Lysozyme in vacuum under the effect of an
  electric field, Examensarbete, Biomedical Center, Uppsala 2011.
  $<$\url{http://www.diva-portal.org/smash/get/diva2:410123/FULLTEXT01.pdf}$>$.

\bibitem{Lee:PR104:254}
T.~D. Lee and C.~N. Yang,  Phys.\ Rev.  \textbf{104} (1), 254--258 (1956).
  $<$\url{http://dx.doi.org/10.1103/PhysRev.104.254}$>$.

\bibitem{Wu:PR105:1413}
C.~S. Wu, E. Ambler, R.~W. Hayward, D.~D. Hoppes, and R.~P. Hudson,  Phys.\
  Rev.  \textbf{105} (4), 1413--1415 (1957).
  $<$\url{http://dx.doi.org/10.1103/PhysRev.105.1413}$>$.

\bibitem{Darquie:Chirality22:870}
B. Darqui{\'e}, C. Stoeffler, A. Shelkovnikov, C. Daussy, A. Amy-Klein, C.
  Chardonnet, S. Zrig, L. Guy, J. Crassous, P. Soulard, P. Asselin, T.~R. Huet,
  {\em{et~al.}},  Chirality  \textbf{22} (10), 870--884 (2010).
  $<$\url{http://doi.wiley.com/10.1002/chir.20911}$>$.

\bibitem{Quack:ProgressTheoChemPhys26:47}
M. Quack, in \emph{Quantum Systems in Chemistry and Physics}, edited by K.
  Nishikawa, J. Maruani, E.~J. Br{\"a}ndas, G. Delgado-Barrio, and P. Piecuch,
  Chap.~3  (Springer Netherlands, Dordrecht, 2012), pp. 47--76.
  $<$\url{http://link.springer.com/10.1007/978-94-007-5297-9}$>$.

\bibitem{Quack:ACIE41:4618}
M. Quack,  Angew.\ Chem.\ Int.\ Ed.  \textbf{41}, 4618--4630 (2002).

\bibitem{Daussy:PRL83:1554}
C. Daussy, T. Marrel, A. Amy-Klein, C.~T. Nguyen, C.~J. Bord\'e, and C.
  Chardonnet,  Phys.\ Rev.\ Lett.  \textbf{83}, 1554--1557 (1999).

\bibitem{Medcraft:ACIE53:11656}
C. Medcraft, R. Wolf, and M. Schnell,  Angew.\ Chem.\ Int.\ Ed.  \textbf{53},
  11656 (2014).

\bibitem{Patterson:Nature497:475}
D. Patterson, M. Schnell, and J.~M. Doyle,  Nature  \textbf{497} (7450),
  475--477 (2013).

\bibitem{Jacob:JCP137:044313}
A. Jacob and K. Hornberger,  J.\ Chem.\ Phys.  \textbf{137} (4), 044313 (2012).
   $<$\url{http://scitation.aip.org/content/aip/journal/jcp/137/4/10.1063/1.4738753}$>$.

\bibitem{Haertelt:JCP128:224313}
M. H{\"a}rtelt and B. Friedrich,  J.\ Chem.\ Phys.  \textbf{128} (22), 224313
  (2008).

\bibitem{Omiste:PCCP13:18815}
J.~J. Omiste, M. Gaerttner, P. Schmelcher, R. Gonz{\'a}lez-F{\'e}rez, L.
  Holmegaard, J.~H. Nielsen, H. Stapelfeldt, and J. K{\"u}pper,  Phys.\ Chem.\
  Chem.\ Phys.  \textbf{13} (42), 18815--18824 (2011).

\bibitem{Nielsen:PRL108:193001}
J.~H. Nielsen, H. Stapelfeldt, J. K{\"u}pper, B. Friedrich, J.~J. Omiste, and
  R. Gonz{\'a}lez-F{\'e}rez,  Phys.\ Rev.\ Lett.  \textbf{108} (19), 193001
  (2012).  $<$\url{http://prl.aps.org/abstract/PRL/v108/i19/e193001}$>$.

\bibitem{Hansen:JCP139:234313}
J.~L. Hansen, J.~J. Omiste~Romero, J.~H. Nielsen, D. Pentlehner, J. Küpper, R.
  Gonz{\'a}lez-F{\'e}rez, and H. Stapelfeldt,  J.\ Chem.\ Phys.  \textbf{139},
  234313 (2013).

\bibitem{Kirste:Science338:1060}
M. Kirste, X. Wang, H.~C. Schewe, G. Meijer, K. Liu, A. van~der Avoird,
  L.~M.~C. Janssen, K.~B. Gubbels, G.~C. Groenenboom, and S.~Y.~T. van~de
  Meerakker,  Science  \textbf{338} (6110), 1060--1063 (2012).
  $<$\url{http://www.sciencemag.org/cgi/doi/10.1126/science.1229549}$>$.

\bibitem{Henson:Science338:234}
A.~B. Henson, S. Gersten, Y. Shagam, J. Narevicius, and E. Narevicius,  Science
   \textbf{338} (6104), 234--238 (2012).
  $<$\url{http://www.sciencemag.org/cgi/doi/10.1126/science.1229141}$>$.

\bibitem{Jankunas:JPCA118:3875}
J. Jankunas, B. Bertsche, and A. Osterwalder,  J.\ Phys.\ Chem.\ A
  \textbf{118} (22), 3875--3879 (2014).

\bibitem{Yang:ARPC58:433}
X. Yang,  Annu.\ Rev.\ Phys.\ Chem.  \textbf{58} (1), 433--459 (2007).
  $<$\url{http://dx.doi.org/10.1146/annurev.physchem.58.032806.104632}$>$.

\bibitem{Chefdeville:Science341:1094}
S. Chefdeville, Y. Kalugina, S.~Y.~T. van~de Meerakker, C. Naulin, F. Lique,
  and M. Costes,  Science  \textbf{341} (6150), 1094--1096 (2013).
  $<$\url{http://www.sciencemag.org/cgi/doi/10.1126/science.1241395}$>$.

\bibitem{Brooks:Science193:11}
P.~R. Brooks,  Science  \textbf{193} (4247), 11 (1976).

\bibitem{Miranda:NatPhys7:502}
M.~H.~G. de~Miranda, A. Chotia, B. Neyenhuis, D. Wang, G. Qu{\'e}m{\'e}ner, S.
  Ospelkaus, J.~L. Bohn, J. Ye, and D.~S. Jin,  Nature Physics  \textbf{7},
  502–507 (2011).
  $<$\url{http://www.nature.com/nphys/journal/vaop/ncurrent/full/nphys1939.html}$>$.

\bibitem{Whittle:JCP22:1943}
E. Whittle, D.~A. Dows, and G.~C. Pimentel,  J.\ Chem.\ Phys.  \textbf{22}
  (11), 1943--1943 (1954).  $<$\url{http://dx.doi.org/10.1063/1.1739957}$>$.

\bibitem{Bondybey:ChemRev96:2113}
V.~E. Bondybey, A.~M. Smith, and J. Agreiter,  Chem.\ Rev.  \textbf{96} (6),
  2113--2134 (1996).

\bibitem{Turgeon:PRA86:062710}
P.~A. Turgeon, P. Ayotte, E. Lisitsin, Y. Meir, T. Kravchuk, and G.
  Alexandrowicz,  Phys.\ Rev.\ A  \textbf{86} (6), 062710 (2012).
  $<$\url{http://link.aps.org/doi/10.1103/PhysRevA.86.062710}$>$.

\bibitem{Sliter:JPCA115:9682}
R. Sliter, M. Gish, and A.~F. Vilesov,  J.\ Phys.\ Chem.\ A  \textbf{115} (34),
  9682--9688 (2011).  $<$\url{http://dx.doi.org/10.1021/jp201125k}$>$.

\end{thebibliography}
\bibliographystyle{tRPC}

\end{document}